\documentclass[journal]{IEEEtran}
\usepackage[noadjust]{cite}
\usepackage{amsmath,amssymb,amsfonts}
\usepackage{graphicx}
\usepackage{textcomp}
\usepackage[caption=false]{subfig}
\usepackage{tabularx}
\def\BibTeX{{\rm B\kern-.05em{\sc i\kern-.025em b}\kern-.08em
    T\kern-.1667em\lower.7ex\hbox{E}\kern-.125emX}}

\begin{document}
\title{Serrodyne frequency translation using time-modulated metasurfaces}
\author{Zhanni Wu, \IEEEmembership{Student Member, IEEE}, Anthony Grbic, \IEEEmembership{Fellow, IEEE}
\thanks{The authors are with Department of Electrical Engineering and Computer Science, University of Michigan, Ann Arbor, Michigan 48109-2122, USA (e-mail: zhanni@umich.edu; agrbic@umich.edu).}
}

\maketitle

\begin{abstract}
Temporally modulated metamaterials have attracted significant attention recently due to their non-reciprocal and frequency converting properties. Here, a transparent, time-modulated metasurface that functions as a serrodyne frequency translator, is reported at X-band frequencies. With a simple biasing architecture, the metasurface provides electrically-tunable transmission phase that covers $360^\circ$. A sawtooth waveform is used to modulate the metasurface, allowing Doppler-like frequency translation. Two such metasurfaces can be cascaded together to achieve magnetless devices that perform either phase or amplitude non-reciprocity.
\end{abstract}


\section{Introduction}
\label{sec:introduction}
\IEEEPARstart{T}{ime-varying} metasurfaces, can exhibit dynamically controllable scattering properties, and allow the real-time manipulation of electromagnetic waves. In addition to providing reconfigurable static functionalities, temporal modulation allows harmonic generation and non-reciprocal responses \cite{sounas2017non,caloz2018electromagnetic}. For example, one can achieve frequency translation, the process of converting an input signal from one frequency to an other, without generating undesired sidebands. In general, there are different ways to achieve frequency translation, including traveling-wave parametric amplifiers \cite{zucker1961traveling} and acousto-optic modulation \cite{risk1984acousto, cheng1992baseband}. The serrodyne frequency translator \cite{jaffe1965microwave, cumming1957serrodyne}, which employs a continuous linear variation of phase (temporal modulation), closely approaches the performance of an ideal frequency translator. Just like a blazed grating (a linear phase gradient in space) converts one spatial frequency to another, the sawtooth phase variation in time (a linear phase gradient in time) converts one temporal frequency to another. In contrast to a sinusoidally driven mixer, a serrodyne frequency translator suppresses the image frequency without added image rejection circuitry. In earlier works, the serrodyne frequency translator has been developed using different phase-tunable devices including ferrite phase shifters \cite{klein1967digilator}, electro-optic waveguides \cite{wong1982electro, johnson2010broadband, houtz2009wideband} and CPW MMIC phase shifters \cite{lucyszyn199424}.

Here, we present a transparent, time-varying metasurface that achieves free-space, serrodyne frequency translation at X-band frequencies. The metasurface provides full phase variation despite its small physical thickness of $9.56$ mm ($0.3\lambda$). Tunable varactors on the metasurfaces, controlled by a simple biasing network, allow electrically-driven, time-modulated surface properties. The proposed metasurface-based, frequency translators may find a wide range of applications including frequency comb generation \cite{wu2010generation, jessen1992generation}, arbitrary waveform generation \cite{cundiff2010optical}, Doppler cloaks \cite{ramaccia2017doppler}, modulated antennas \cite{taravati2017mixer, henthorn2017bit, zhang2018space, zhao2018programmable}, radar pulse-shaping \cite{bekele2013pulse, ghelfi2012photonic}, and magnetless nonreciprocal devices \cite{shaltout2015time, hadad2015space, shi2016dynamic, taravati2017nonreciprocal, estep2014magnetic, sounas2013giant, qin2014nonreciprocal, ramaccia2015nonreciprocal}. 

\begin{figure}[t]
\centering
\includegraphics[clip,width=1\columnwidth]{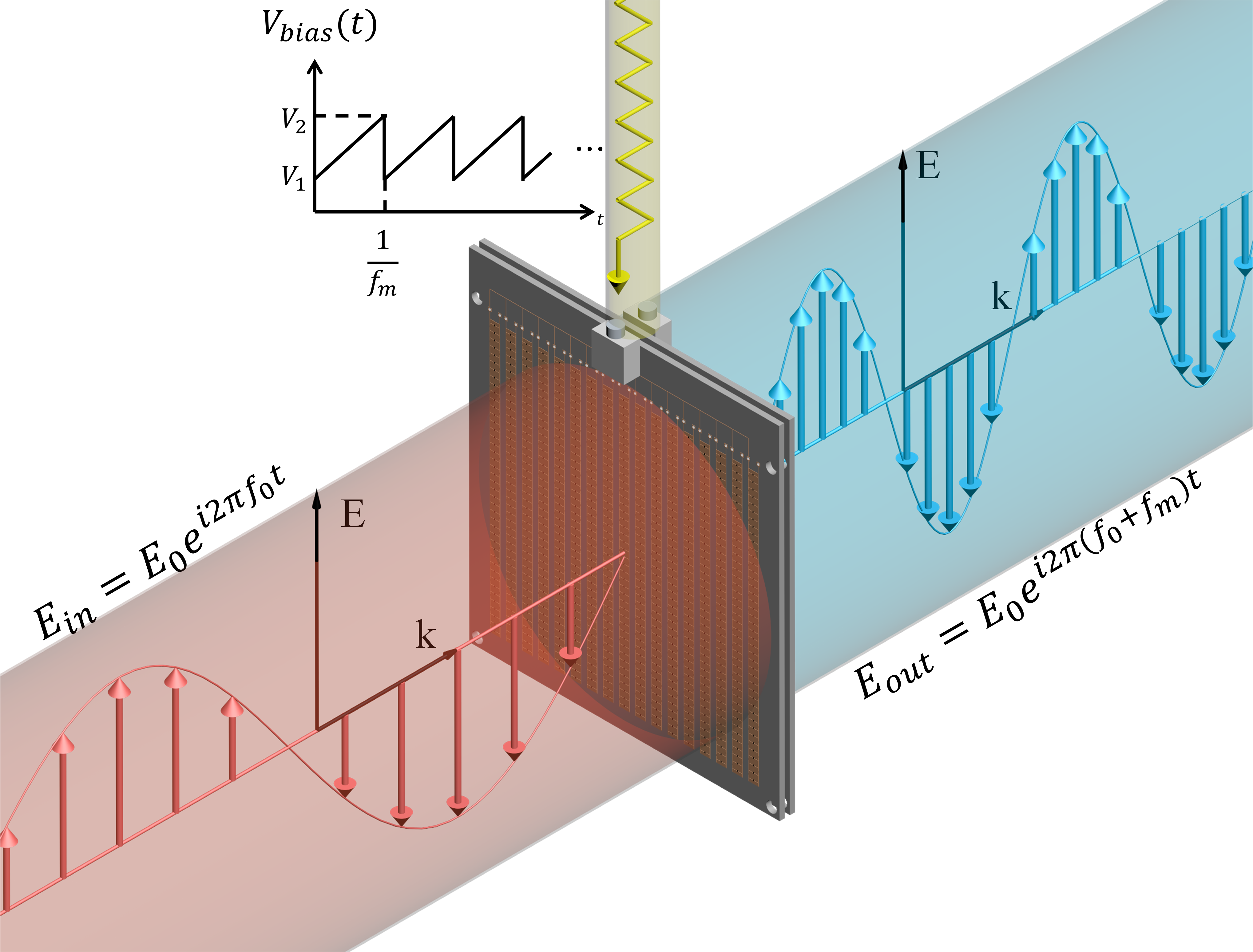}% Here is how to import EPS art
\caption{\label{fig:scheme} A transparent, metasurface-based serrodyne frequency translator.}
\end{figure}

\section{Design and simulation of a transparent, serrodyne frequency translator.}
In this section, we propose and describe the corresponding circuit design and metasurface realization for the serrodyne frequency translator. Furthermore, analytical as well as harmonic balance simulation of the transmission spectrum frequency translator are reported.

\subsection{Circuit design and metasurface realization}
Recently \cite{PhysRevX.9.011036}, we demonstrated a transparent, tunable metasurface that functions as a bandpass phase shifter. The metasurface is impedance-matched to free space and exhibits an electrically-tunable electrical thickness. When modulated with a sawtooth waveform, $V_{bias}(t)$, with a repetition rate of $f_m$, the metasurface provides a periodic transmission phase $\phi(t)$ and transmission amplitude $A(t)$. As shown in Fig. \ref{fig:scheme}, for an incident signal of the form $E_{in}=E_0e^{i2\pi f_0t}$, the transmitted signal may be expressed as
\begin{equation}\label{TransSignal_t}
E_{out}=E_0e^{i2\pi f_0t}A(t)e^{j\phi(t)},
\end{equation}
assuming that the modulation frequency is much lower than the signal frequency: $f_m\ll f_0$. Since the transmission coefficient $h(t)=A(t)e^{i\phi(t)}$ is periodic in time, the spectrum of the transmitted wave can be written as
\begin{equation}\label{TransSignal_Spec}
E_{out}(\omega)=E_0\sum_{n=-\infty}^{\infty}a_n\delta(\omega-\omega_0-2\pi nf_m),
\end{equation}
where $a_n = 2\pi f_m\int_{-\frac{1}{2f_m}}^{\frac{1}{2f_m}}h(t)e^{-i2\pi nf_mt}$. 
Clearly when $A(t)=1$ is unity and $\phi(t)=f_mt$ is linear with time, the phase shifter functions as an ideal frequency translator:
\begin{equation}\label{FTranslation}
E_{out}(\omega)=2\pi\delta(\omega-\omega_0-2\pi f_m)
\end{equation}

\begin{figure}[t]
\centering
\subfloat[\label{sfig:BFSquartercircuit}]{%
  \includegraphics[clip,width=0.7\columnwidth]{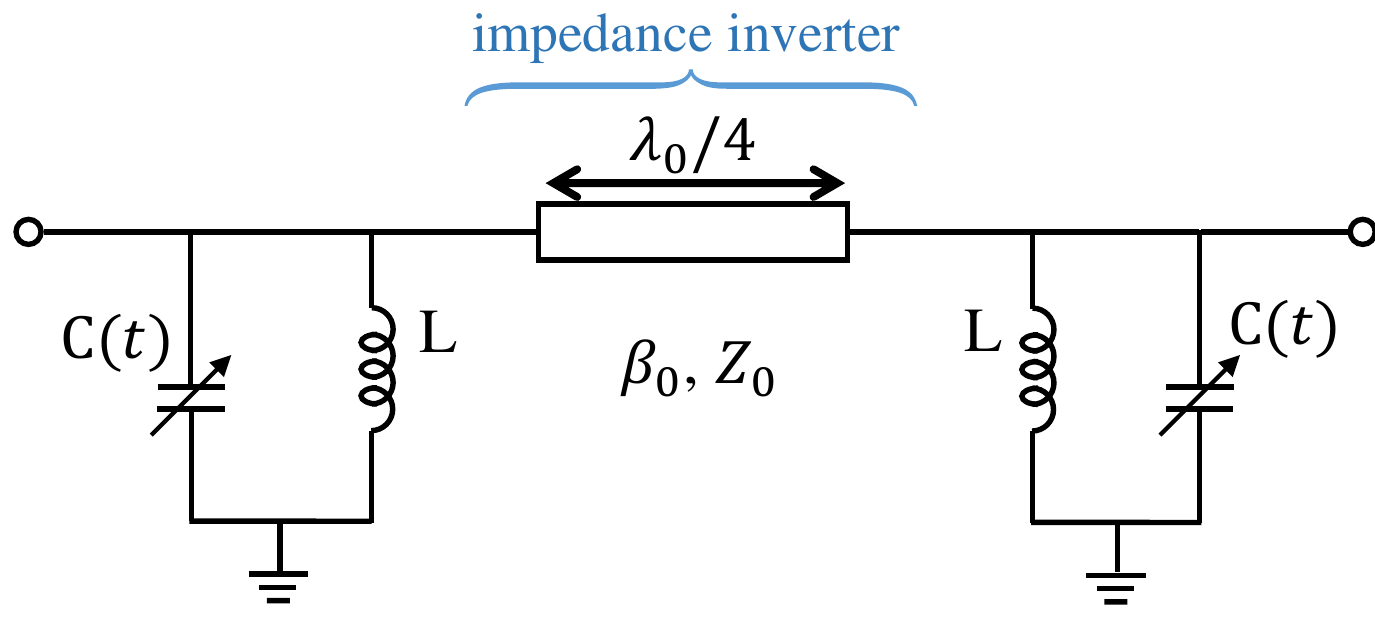}%
}\qquad
\subfloat[\label{sfig:BFSextract}]{%
  \includegraphics[clip,width=1\columnwidth]{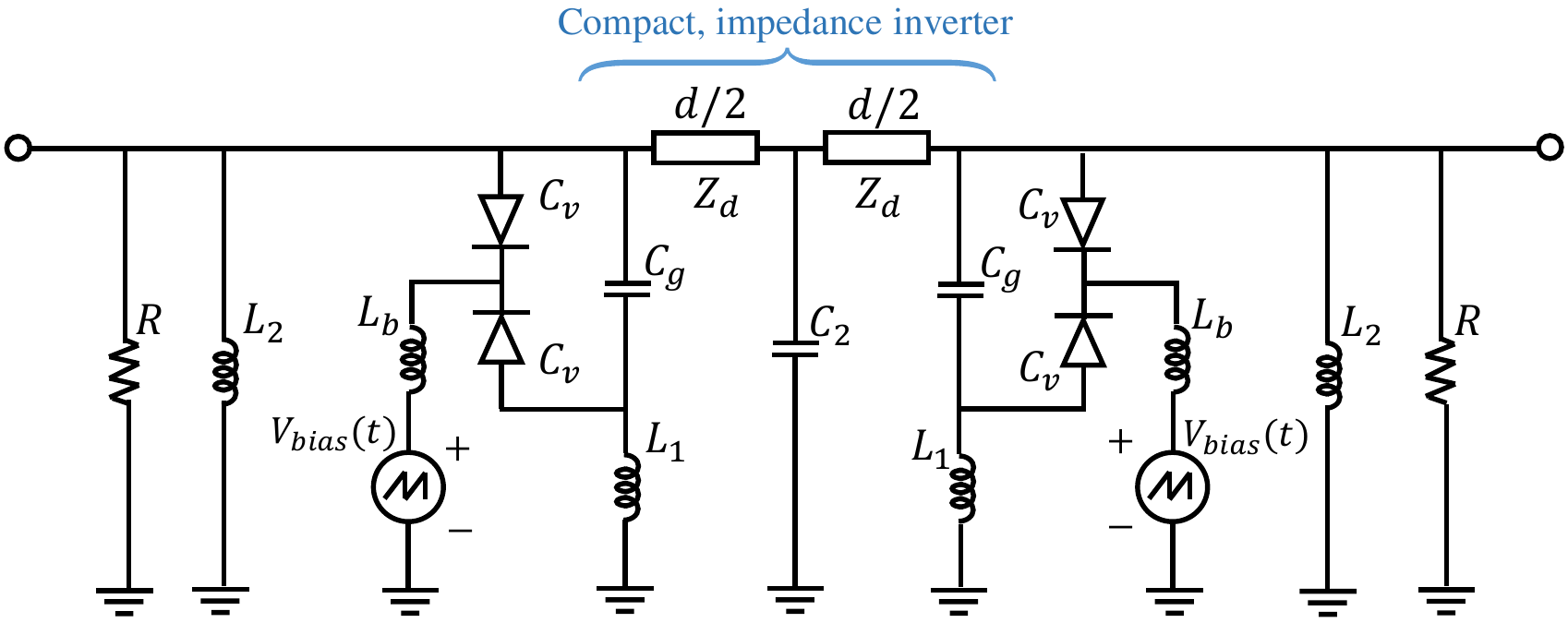}%
}\qquad
\caption{(a) A tunable, transmission-line, phase shifter that maintains an impedance match (low insertion loss) over its tunable range. (b) Extracted circuit model of the designed transparent, tunable metasurface phase shifter.}
\label{fig:BFSCircuit}
\end{figure}

The realization of a reflectionless (transparent) metasurface that functions as a phase shifter, with a time-dependent transmission phase, requires control of both its electric and magnetic surface properties. The ratio of electric to magnetic surface susceptibilities must remain constant and equal to the square of the free-space wave impedance, while their product must linearly increase (or decrease) over each modulation period. This can be achieved easily using the simple bandpass filter topology depicted in Fig. \ref{sfig:BFSquartercircuit}. The bandpass filter contains two shunt resonators (LC tank circuits) separated by a $90^\circ$ phase delay (an impedance inverter). The impedance inverter transforms the second shunt LC resonant circuit to a series LC resonant circuit, allowing both the electric (shunt circuit) and magnetic (series circuit) response to be tuned with variable capacitors $C(t)$ \cite{PhysRevX.9.011036}, controlled by a single bias waveform. The time-dependent capacitors $C(t)$ are realized as varactor diodes. The capacitance $C(t)$ can be written as,
\begin{equation}\label{CapacitanceDef}
C(t)=C_0+\Delta C(t)
\end{equation}
The static capacitance $C_0$ is chosen to resonate with the shunt LC tank at an operating frequency of $f_0=10$ GHz,
\begin{equation}\label{Resonace}
\omega_0 LC_0=1,
\end{equation}
where $\omega_0=2\pi f_0$, $\Delta C(t)$ is a time-modulated capacitance with a tuning range relatively small compared to $C_0$. Close to the operating resonant frequency, the image impedance $Z_i$ and transmission phase of the phase shifter depicted in Fig. \ref{sfig:BFSquartercircuit} can be calculated as \cite{pozar2009microwave}:
\begin{equation}\label{ImageValue}
\begin{aligned}
Z_i&=Z_0\sqrt{\frac{1}{1-Z_0^2\omega_0^2\Delta C^2(t)}}\\
kd&=\frac{\pi}{2}-Z_0\omega_0\Delta C(t)
\end{aligned}
\end{equation}
where $Z_0=377\Omega$ is the free-space impedance. For capacitance variations $Z_0^2\omega_0^2\Delta C^2(t)\ll 1$, the phase shifter remains impedance-matched to free space: $Z_i\approx Z_0$. In addition, the transmission phase of the phase shifter is linearly proportional to the time-varying $C(t)$.

\begin{figure}[t]
\centering
\subfloat[\label{sfig:BFSboard}]{%
  \includegraphics[clip,width=0.6\columnwidth]{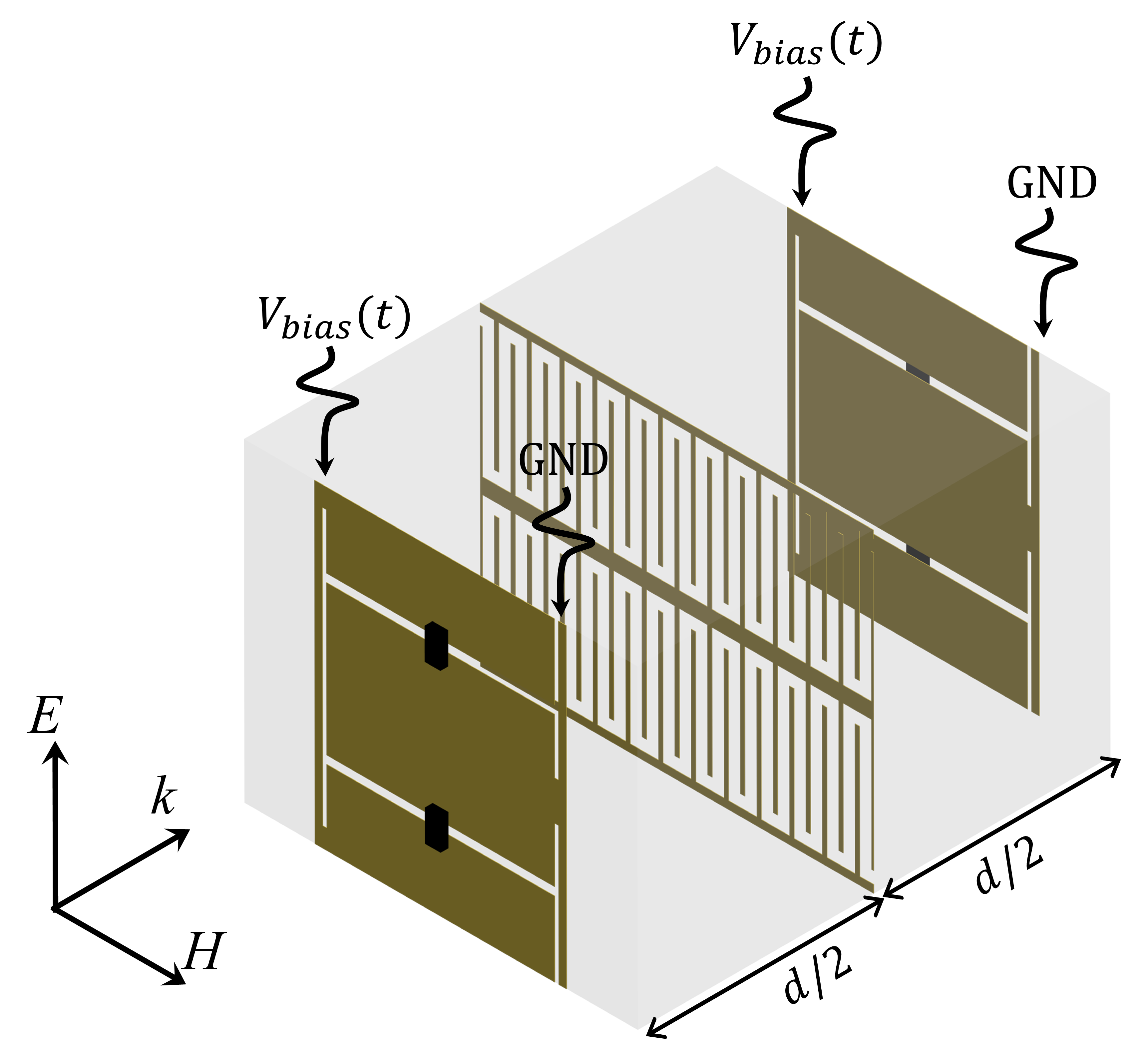}%
}\qquad
\subfloat[\label{sfig:BFSCellout}]{%
  \includegraphics[clip,width=0.44\columnwidth]{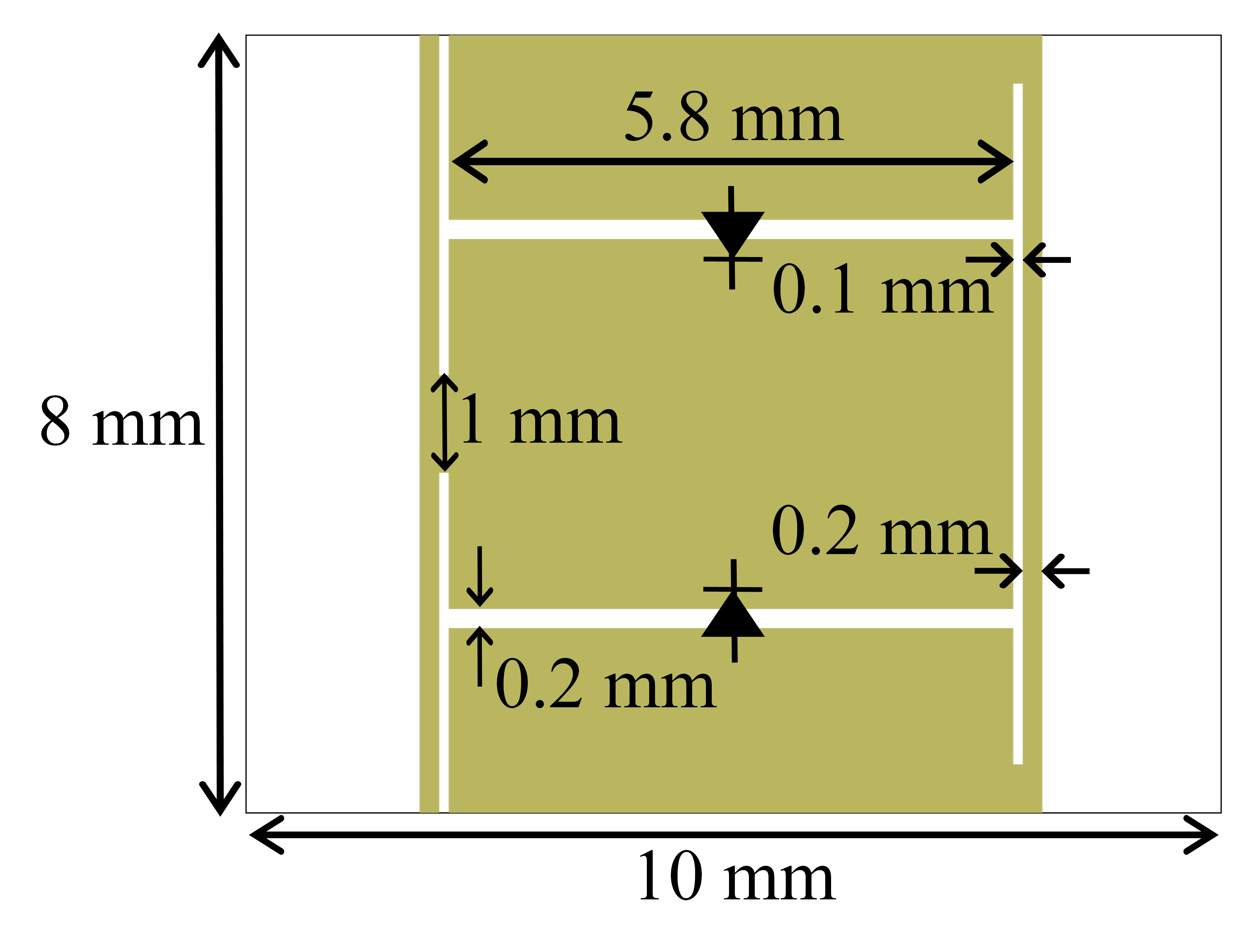}%
}\hfill 
\subfloat[\label{sfig:BFSCellin}]{%
  \includegraphics[clip,width=0.54\columnwidth]{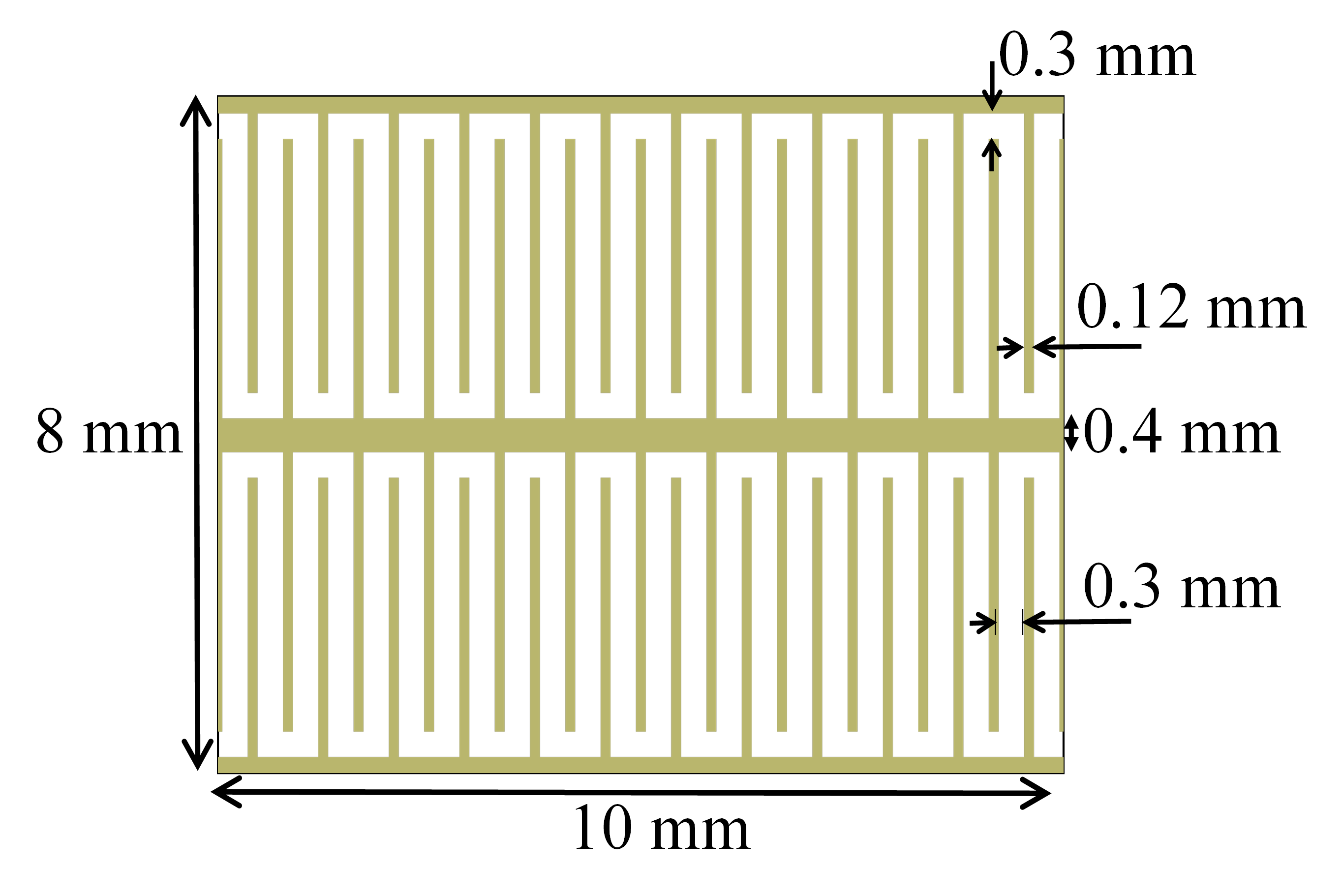}%
}\qquad
\caption{(a) Topology of the transparent, tunable metasurface phase shifter. The metasurface consists of three cascaded patterned metal sheets. The overall thickness is $d=3.15$ mm. (b) Unit cell dimensions for outer sheets. (c) Unit cell dimensions for inner sheet.}
\label{fig:BFSMeta}
\end{figure}

\begin{figure*}
\centering
\includegraphics[clip,width=1.98\columnwidth]{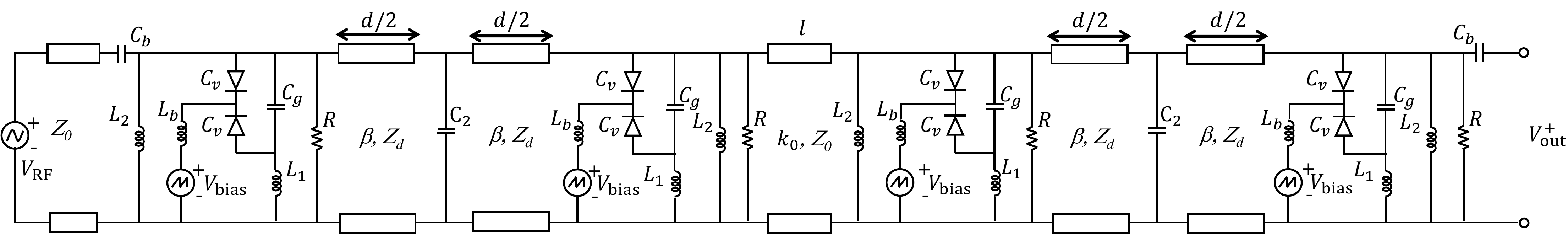}% Here is how to import EPS art
\caption{\label{fig:FreqConverter} Serrodyne frequency translator consisting of two cascaded, transparent tunable metasurface phase shifters. $k_0$ and $\beta$ represent the free-space and substrate wave number respectively.}
\end{figure*}

The phase shifter in Fig. \ref{sfig:BFSquartercircuit} can be realized as an electrically tunable metasurface, shown in Fig. \ref{fig:BFSMeta} \cite{PhysRevX.9.011036}. Varactor diodes (MAVR-000120-1411 from MACOM) are integrated on both sides of the metasurface to achieve the required tunability. A miniaturized impedance inverter was designed to replace the $\lambda_0/4$ spacer and reduce the physical thickness of the metasurface \cite{PhysRevX.9.011036}. The extracted circuit model of the designed metasurface is shown in Fig. \ref{sfig:BFSextract}. In order to achieve $360^\circ$ of phase variation with low transmission amplitude variation, two metasurfaces were cascaded with a free-space spacer between them of length $l$. A circuit model of the resulting cascaded structure is shown in Fig. \ref{fig:FreqConverter}. The values of the extracted circuit parameters are shown in Table \ref{tab:CircuitParam}.

\subsection{Analytical and harmonic balance simulation results of the transmission spectrum}
The circuit was simulated with the commercial circuit solver Keysight Advanced Design System (ADS). The diode $C_v$ was modeled using the spice model of the varactor \cite{macom}.  Note that the values shown in Table \ref{tab:CircuitParam} are cell-averaged values that are scaled by the length-to-width ratio of the unit cell \cite{PhysRevX.9.011036}. The circuit model was verified by showing close agreement between simulations of the circuit model depicted in Fig. \ref{sfig:BFSextract} and full-wave simulations (performed using Ansys HFSS) of the metasurface shown in Fig. \ref{sfig:BFSboard} for various DC bias voltages (See Fig. \ref{fig:BiasCompare}).

\begin{table}[t]%The best place to locate the table environment is directly after its first reference in text
\caption{\label{tab:CircuitParam}%
Values for components shown in Fig. \ref{fig:FreqConverter}.} 
\centering
\setlength{\tabcolsep}{0.5pt}
\begin{tabular}{|>{\centering\arraybackslash}p{0.2\columnwidth}| >{\centering\arraybackslash}p{0.7\columnwidth}|}
\hline
Symbol& 
Detailed information and values\\
\hline
$Z_0$ & Extracted free space impedance, $301.6$ $\Omega$ \\
$Z_d$ & Extracted substrate impedance, $203.34$ $\Omega$ \\
$C_g$ &Extracted pattern capacitance, $0.1$ pF\\
$L_1$&Extracted pattern inductance, $0.4$ nH\\
$L_2$&Extracted pattern inductance, $1.84$ nH\\
$C_2$ &Extracted pattern capacitance, $0.675$ pF\\
$R$&Extracted pattern resistance, $2400$ $\Omega$\\
$C_v$&Varactor diode capacitance\\
$C_b$&Low frequency block, 100 pF\\
$L_b$&High frequency block, 100 nH\\
$\beta d/2$&Substrate thickness, electrical length of $28.03^\circ$\\
$k_0l$&Free-space spacer, electrical length of $40^\circ$\\
\hline
\end{tabular}
\end{table}

\begin{figure}[t]
\centering
\subfloat[\label{sfig:SimVol}]{%
  \includegraphics[clip,width=0.32\columnwidth]{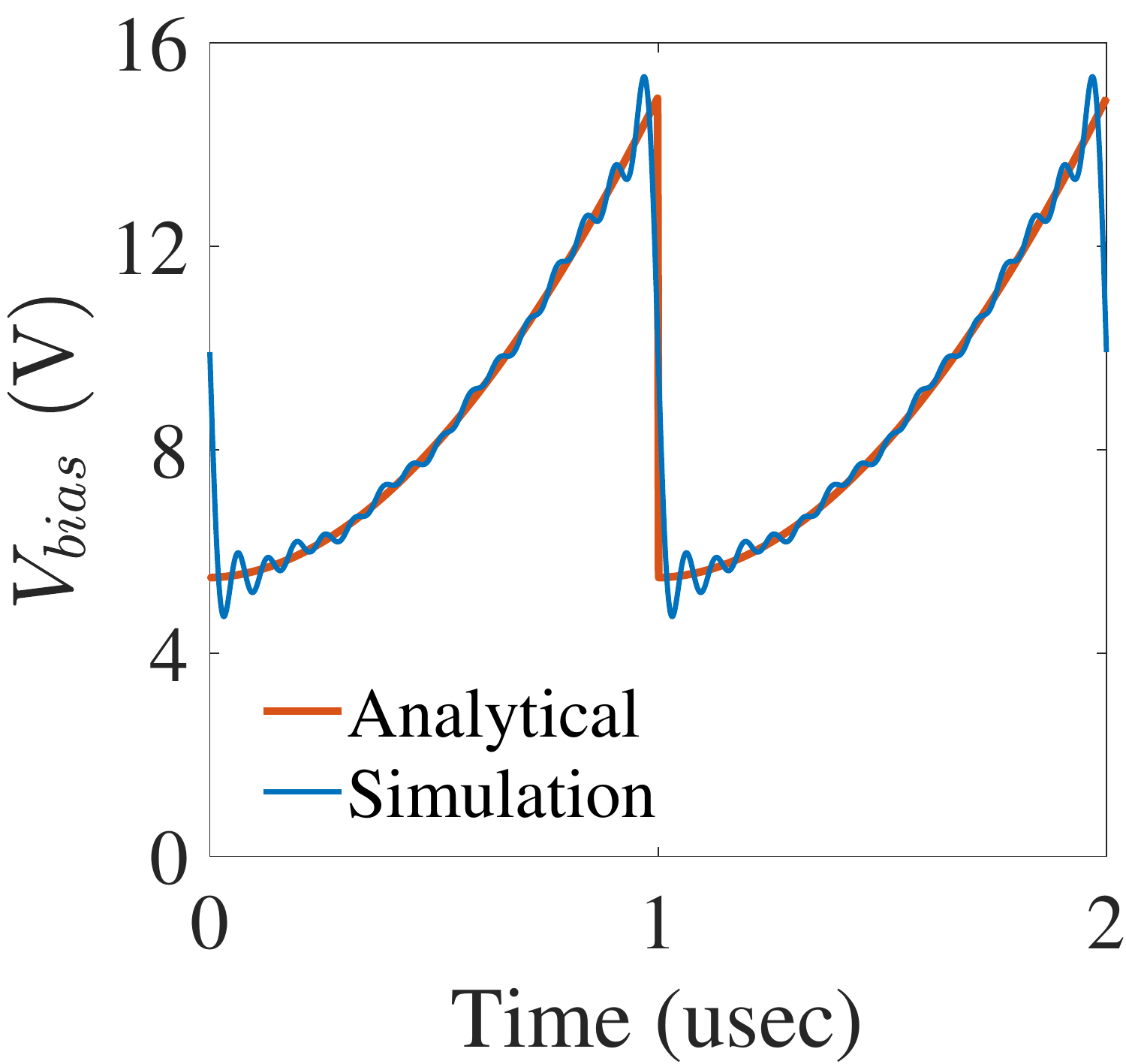}%
}\hfill
\subfloat[\label{sfig:SimA}]{%
  \includegraphics[clip,width=0.32\columnwidth]{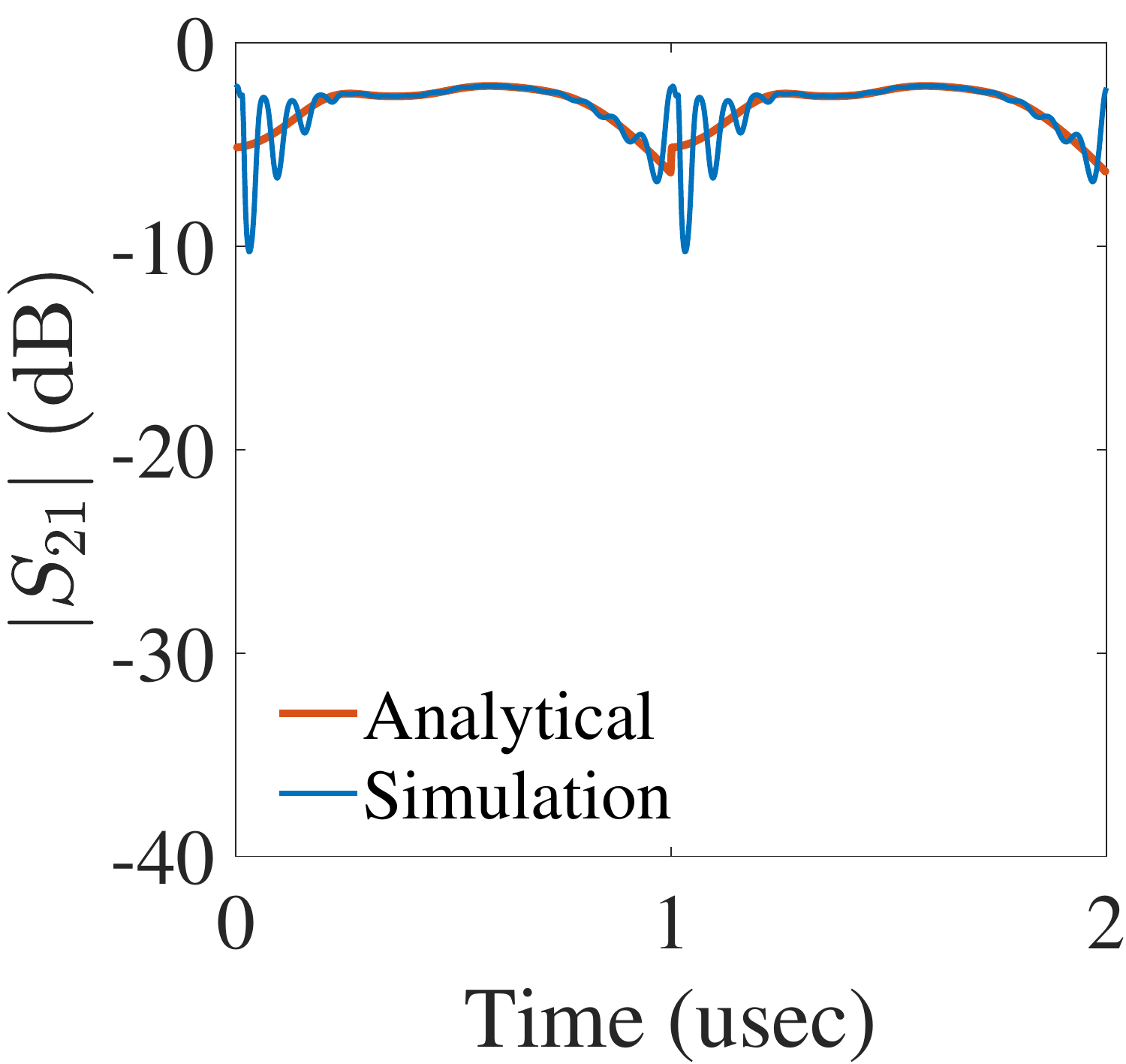}%
}\hfill
\subfloat[\label{sfig:Simphi}]{%
  \includegraphics[clip,width=0.32\columnwidth]{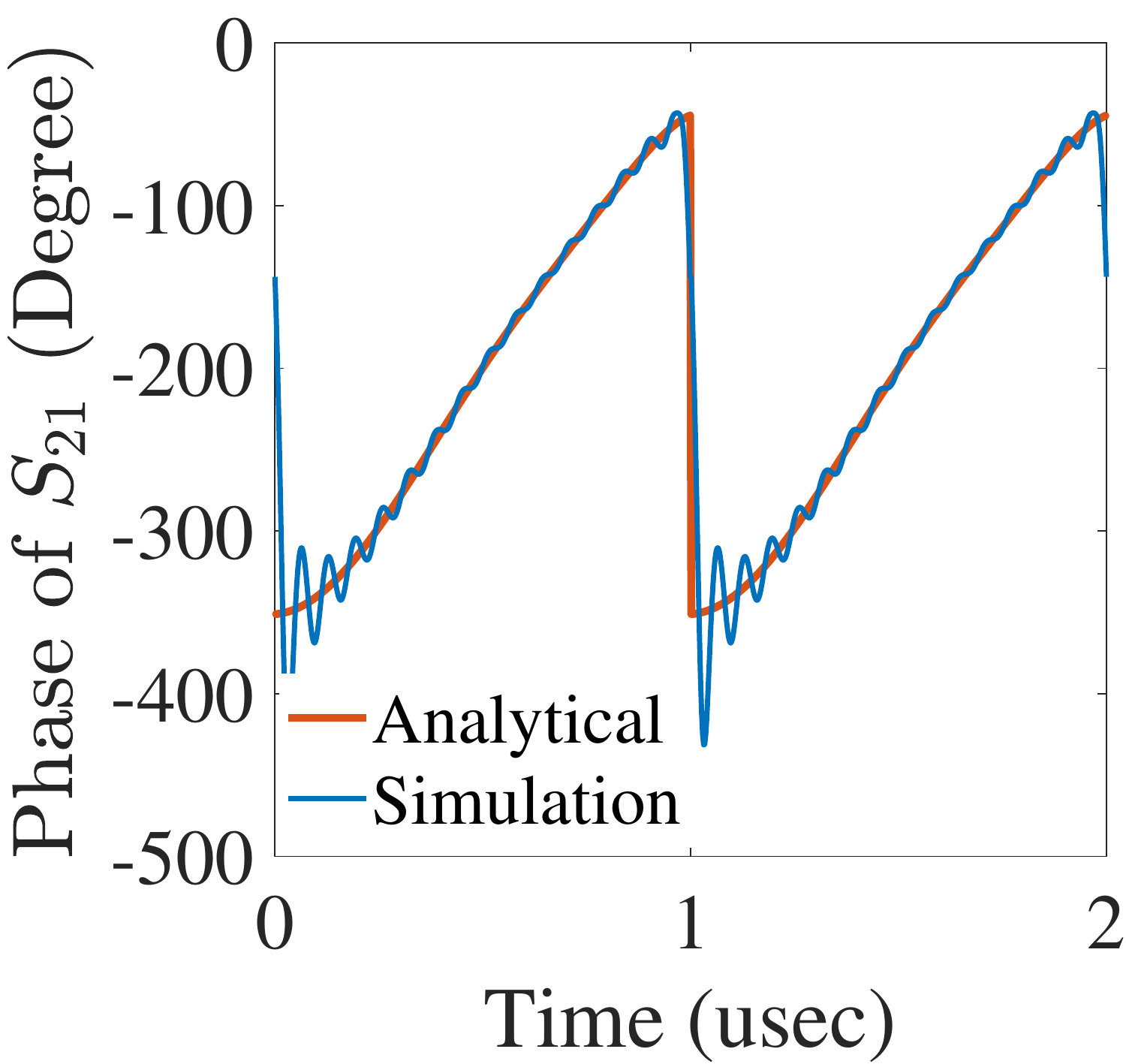}%
}\qquad
\subfloat[\label{sfig:SimSpec_Any}]{%
  \includegraphics[clip,width=0.45\columnwidth]{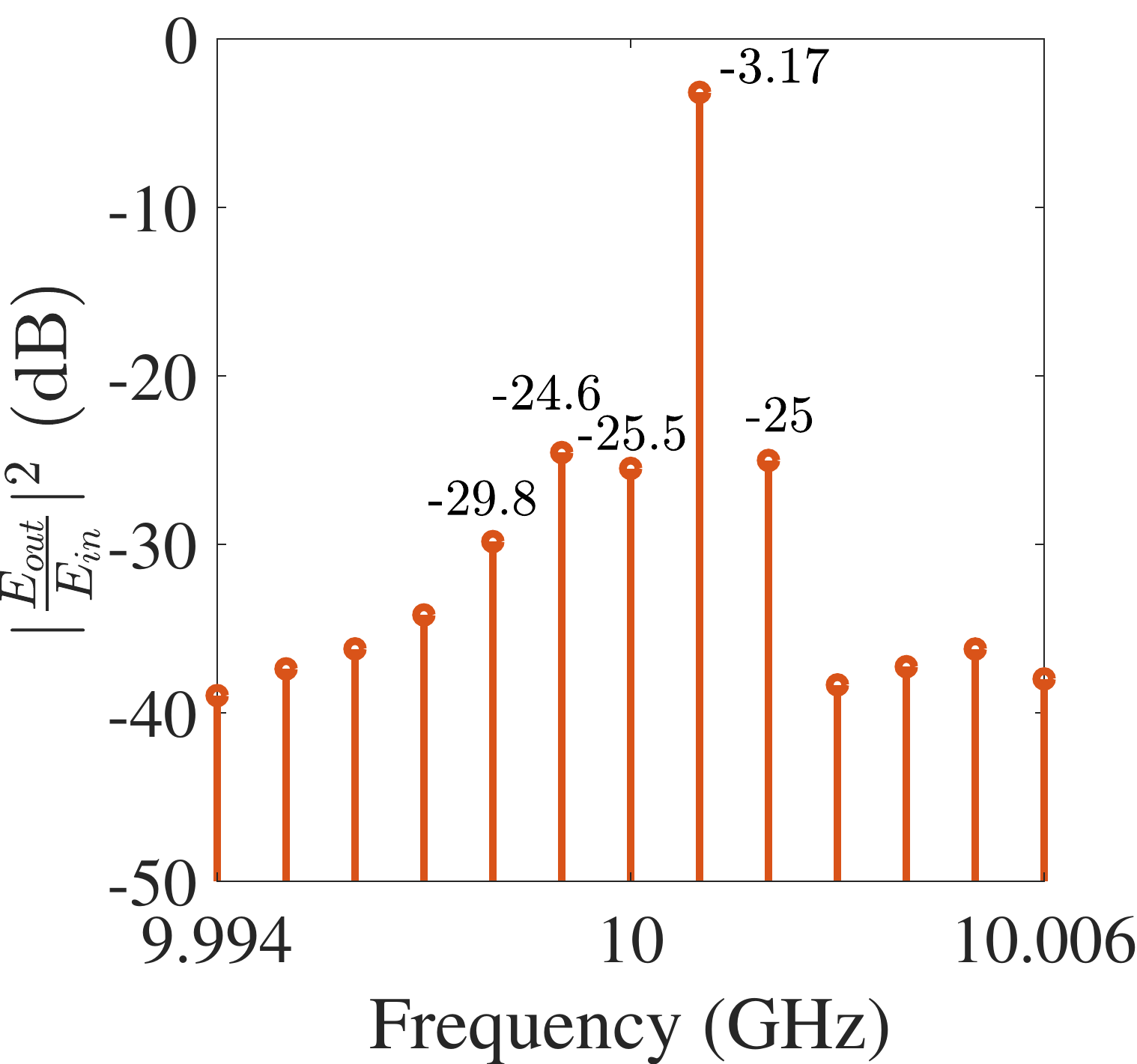}%
}\hfill
\subfloat[\label{sfig:SimSpec_ADS}]{%
  \includegraphics[clip,width=0.45\columnwidth]{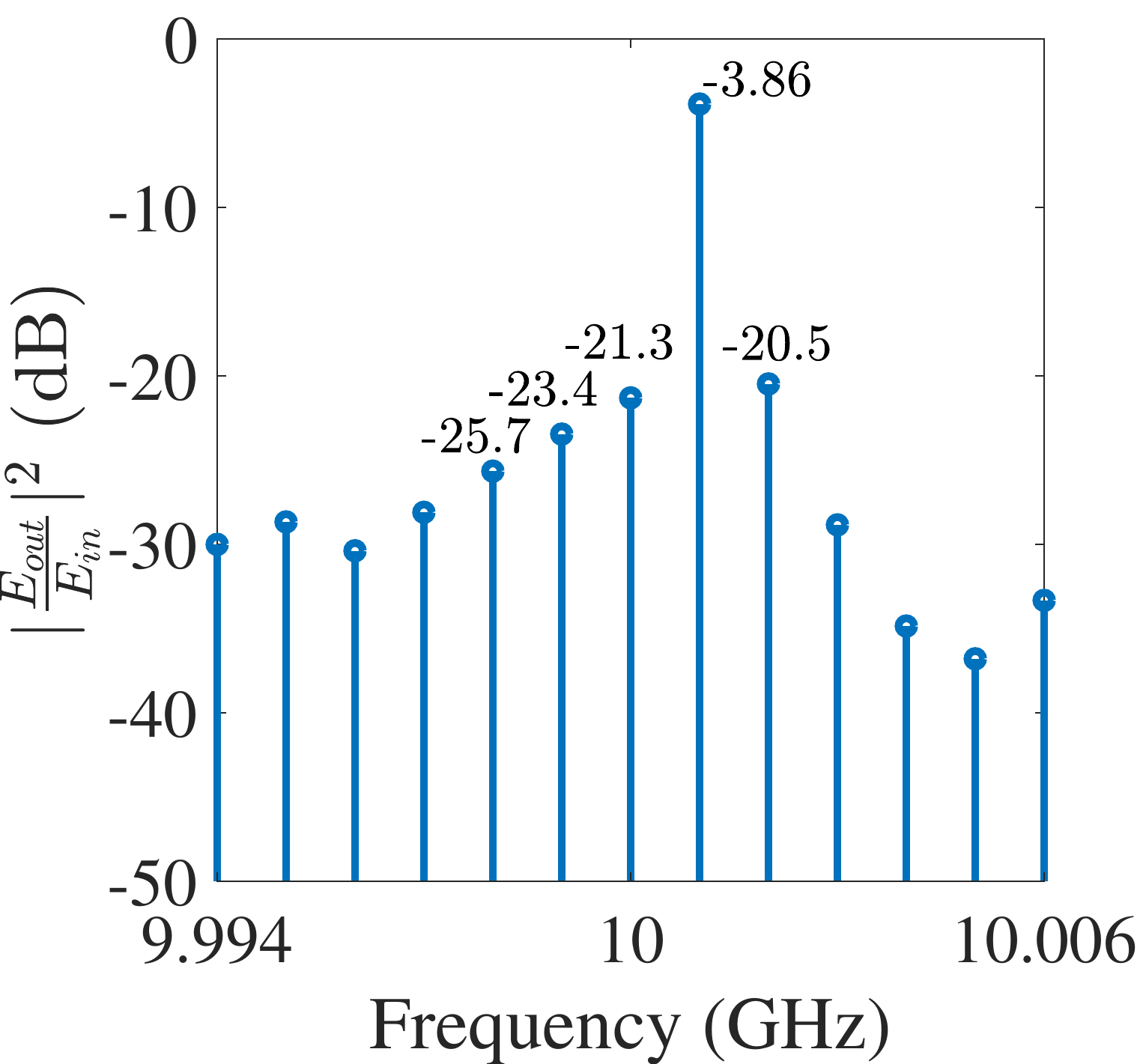}%
}
\caption{(a) Bias waveform used in simulation. (b) Transmission amplitude of the frequency translator with respect to time, at operating frequency of 10 GHz. (c) Transmission phase of the frequency translator with respect to time, at operating frequency of 10 GHz. (d) Analytical transmission spectrum of a 10 GHz signal transmitted through the frequency translator. (e) Simulated transmission spectrum of a 10 GHz signal transmitted through the frequency translator.}
\label{fig:Simulation}
\end{figure}

From the DC simulation of the circuit model shown in Fig. \ref{fig:FreqConverter}, the transmission amplitude and phase versus bias voltage relationships ($A(V_{bias})$, $\phi(V_{bias})$) were extracted. An RF bias waveform with a repetition rate of $f_m=1$ MHz was used to modulate the metasurfaces. Over $1$ $\mu$s modulation period, the bias waveform $V_{bias}(t)$ is modeled as a third-order polynomial with respect to time. Using Eq. \ref{TransSignal_t} and Eq. \ref{TransSignal_Spec}, the transmission spectrum of the frequency translator can be calculated from the extracted $A(V_{bias})$ and $\phi(V_{bias})$. The coefficients of the third-order polynomial function $V_{bias}(t)$ were optimized to provide the highest frequency translation and lowest sidebands. The optimized sawtooth waveform is shown in Fig. \ref{sfig:SimVol} (red line). The corresponding transmission amplitude and phase ($A(t)$, $\phi(t)$) are shown in Fig. \ref{sfig:SimA} and Fig. \ref{sfig:Simphi} respectively (red lines). The incident frequency was set to $f_0 = 10$ GHz. The transmission spectrum calculated using Eq. \ref{TransSignal_t} and Eq. \ref{TransSignal_Spec} is shown in Fig. \ref{sfig:SimSpec_Any}. The spectrum clearly shows a Doppler shift to a frequency of $f_0+f_m=10.001$ GHz. A $3.17$ dB conversion loss and $21.43$ dB sideband suppression is achieved.

A harmonic balance simulation of the circuit model shown in Fig. \ref{fig:FreqConverter} was performed using Keysight ADS. The incident signal was set to an amplitude of $-26.5$ dBm at an operating frequency of $f_0 = 10$ GHz. The optimized bias waveform $V_{bias}(t)$ was approximated with 15 harmonics in simulation, as shown in Fig. \ref{sfig:SimVol} (blue line). The corresponding simulated transmission amplitude and phase ($A(t)$, $\phi(t)$) are shown in Fig. \ref{sfig:SimA} and Fig. \ref{sfig:Simphi} respectively (blue lines). The simulated output spectrum is shown in Fig. \ref{sfig:SimSpec_ADS}. A $3.86$ dB conversion loss and $17.44$ dB sideband suppression is achieved in simulation.

\section{Fabrication and measurement of the transparent, serrodyne frequency translator}
\begin{figure}[t]
\centering\includegraphics[clip,width=0.9\columnwidth]{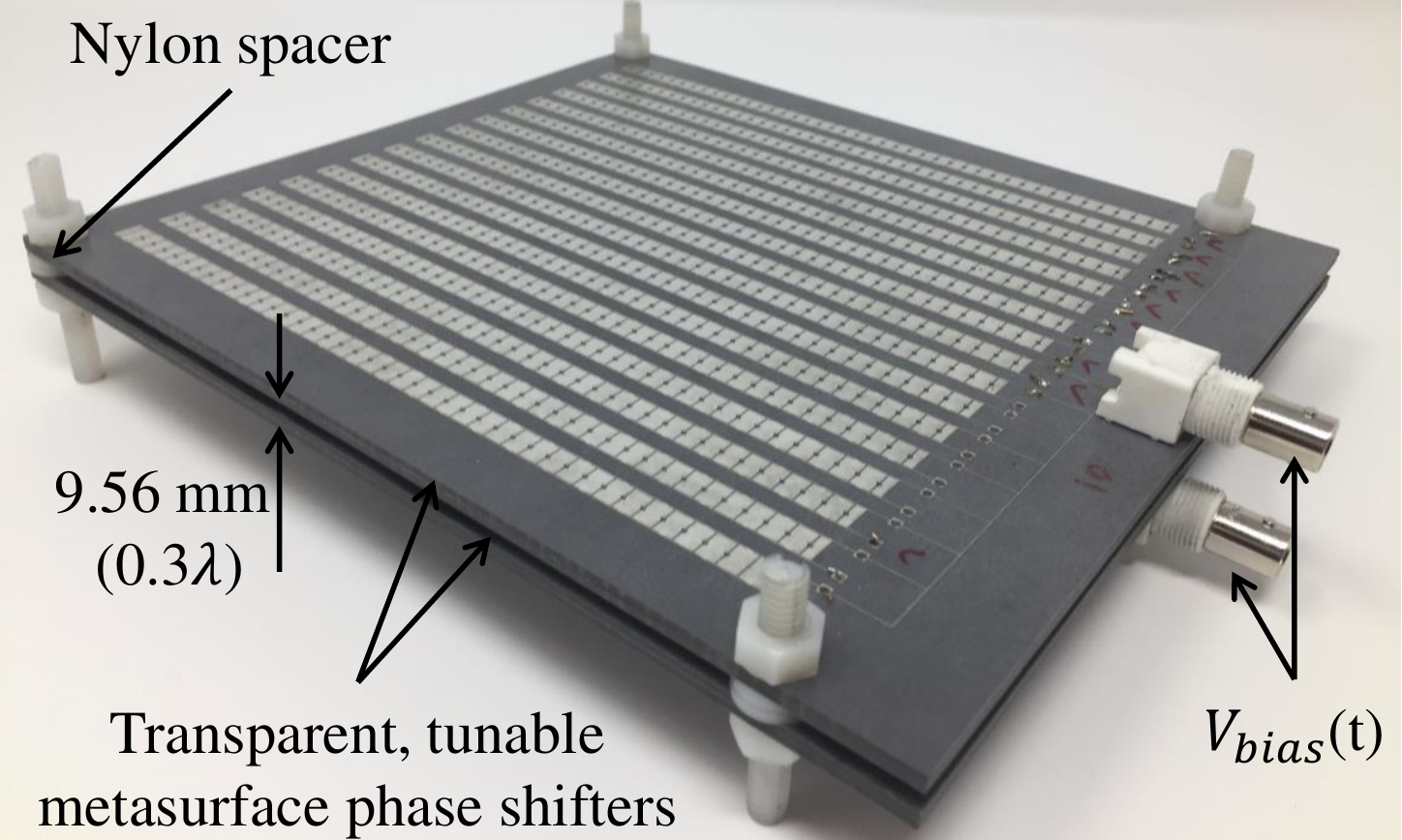}% Here is how to import EPS art
\caption{\label{fig:Fab} Photograph  of the fabricated serrodyne frequency translator.}
\end{figure}

Two of the phase shifting metasurfaces were fabricated \cite{PhysRevX.9.011036}. A RT/Duroid 5880 substrate ($\epsilon_r=2.2$ and $\tan{\delta}=0.0009$) was used in the design. Nylon spacers with a thickness of $l=3.26$ mm were used to separate the two metasurfaces, as shown in Fig. \ref{fig:Fab}. The total thickness of the fabricated frequency translator is approximately $9.56$ mm (0.32 $\lambda$). A total of 5712 MAVR-000120-1411 varactor diodes were integrated in the design. The varactor diodes were controlled simultaneously with a single bias signal. The time-modulated bias signal was connected to the metasurfaces through BNC connectors mounted to each board. 

\begin{figure}[t]
\centering
\subfloat[\label{sfig:Amp345}]{%
  \includegraphics[clip,width=0.32\columnwidth]{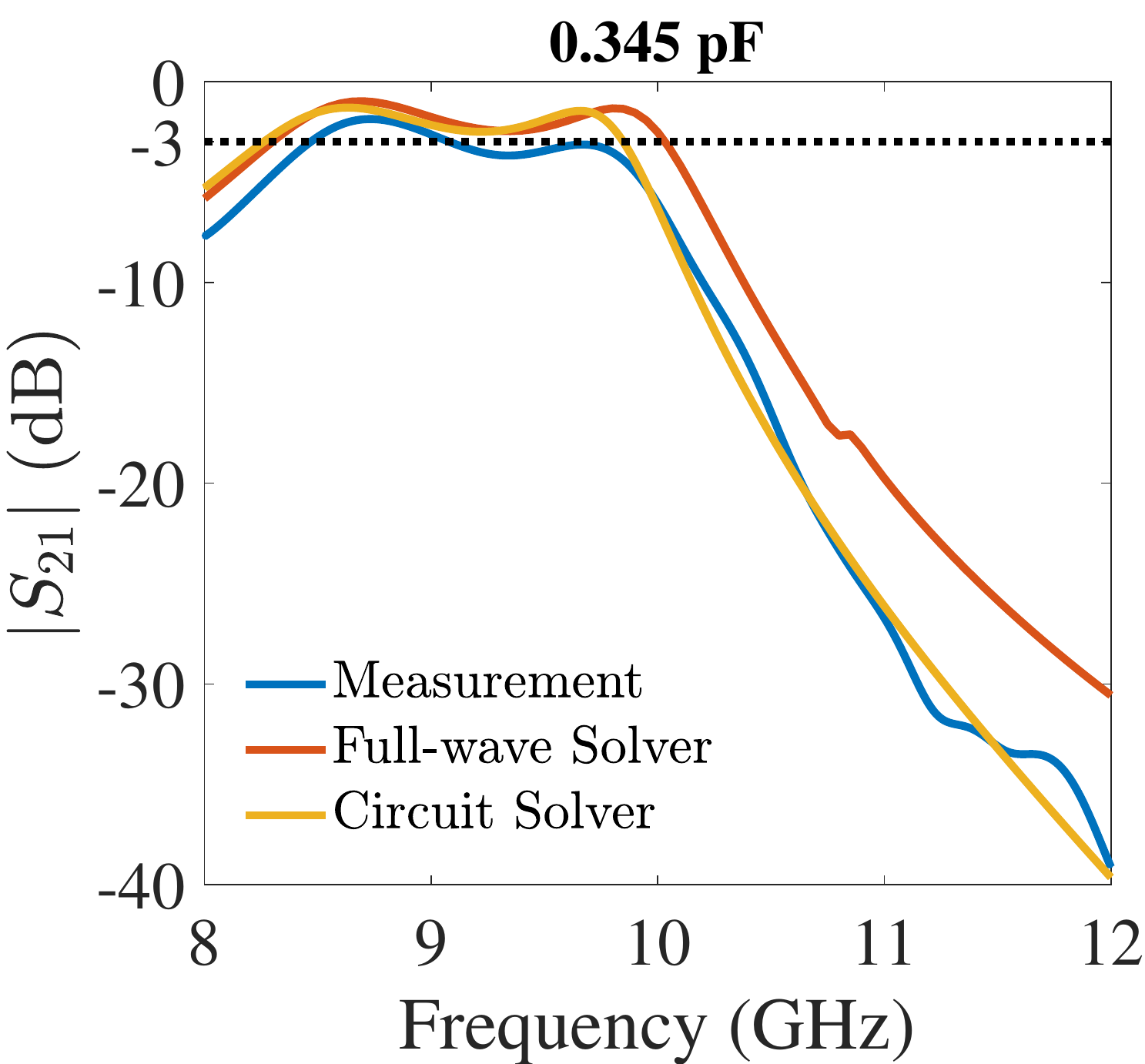}%
}\hfill
\subfloat[\label{sfig:Amp24}]{%
  \includegraphics[clip,width=0.32\columnwidth]{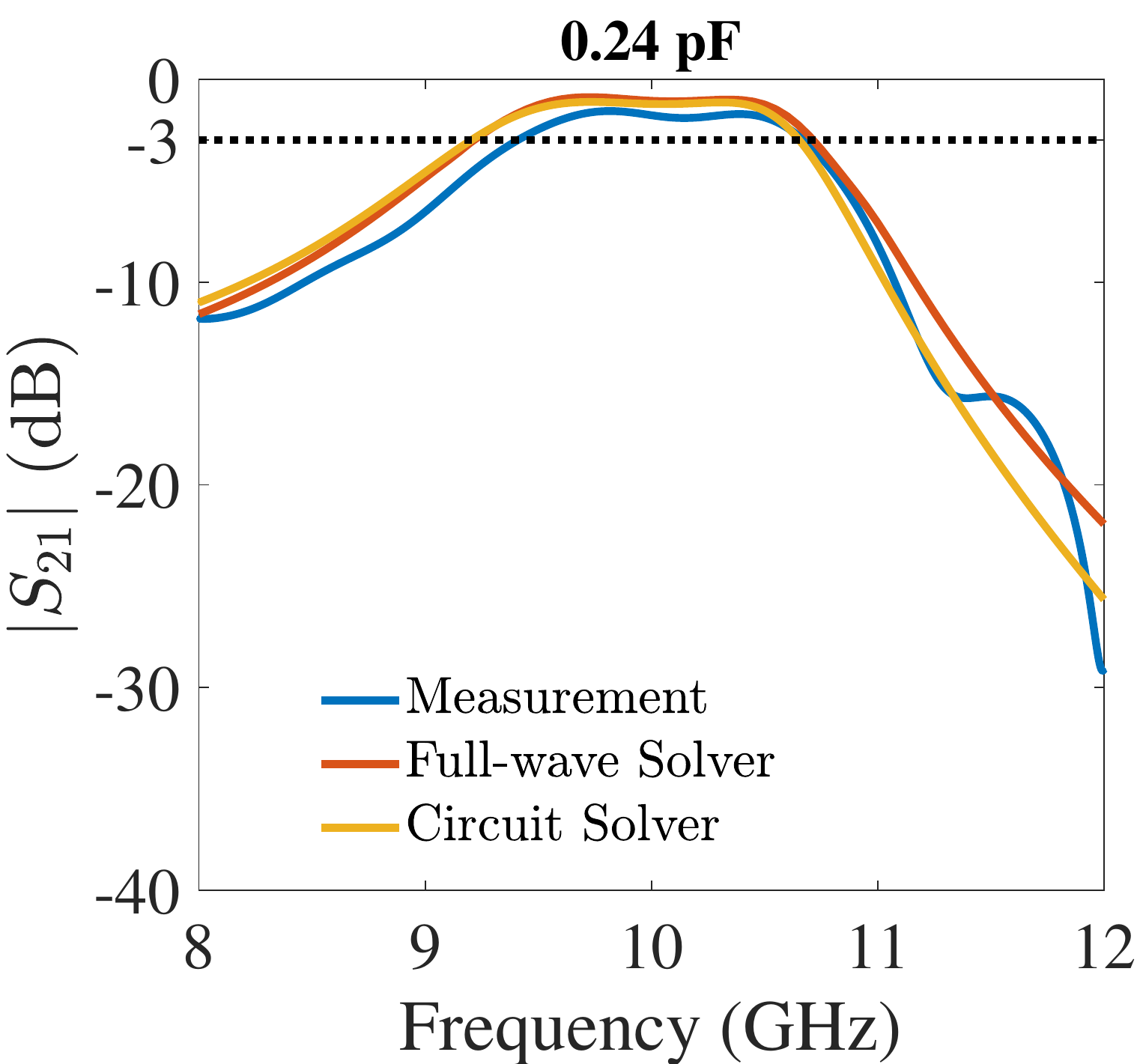}%
}\hfill
\subfloat[\label{sfig:Amp18}]{%
  \includegraphics[clip,width=0.32\columnwidth]{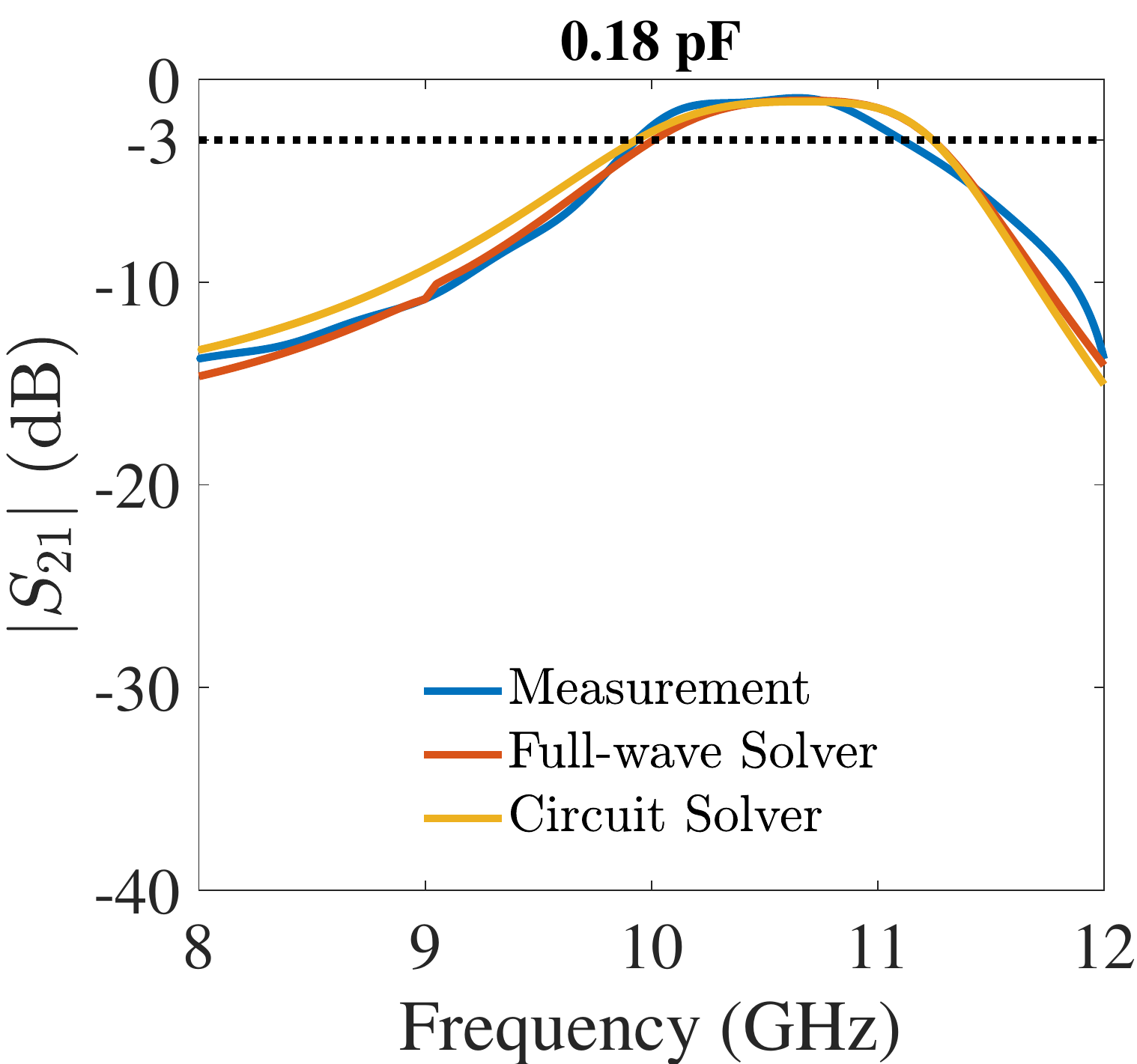}%
}\qquad
\subfloat[\label{sfig:Phase345}]{%
  \includegraphics[clip,width=0.32\columnwidth]{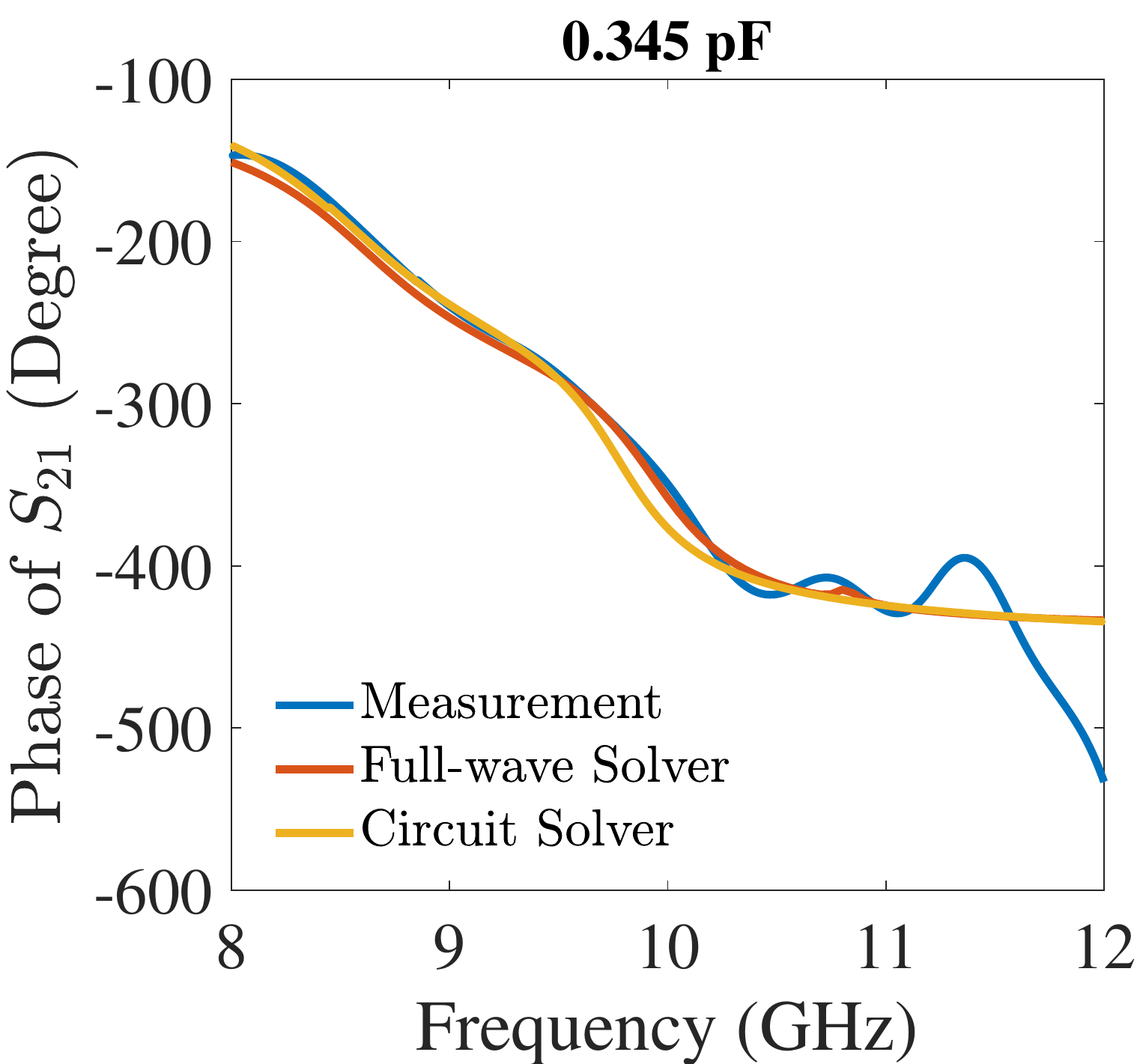}%
}\hfill
\subfloat[\label{sfig:Phase24}]{%
  \includegraphics[clip,width=0.32\columnwidth]{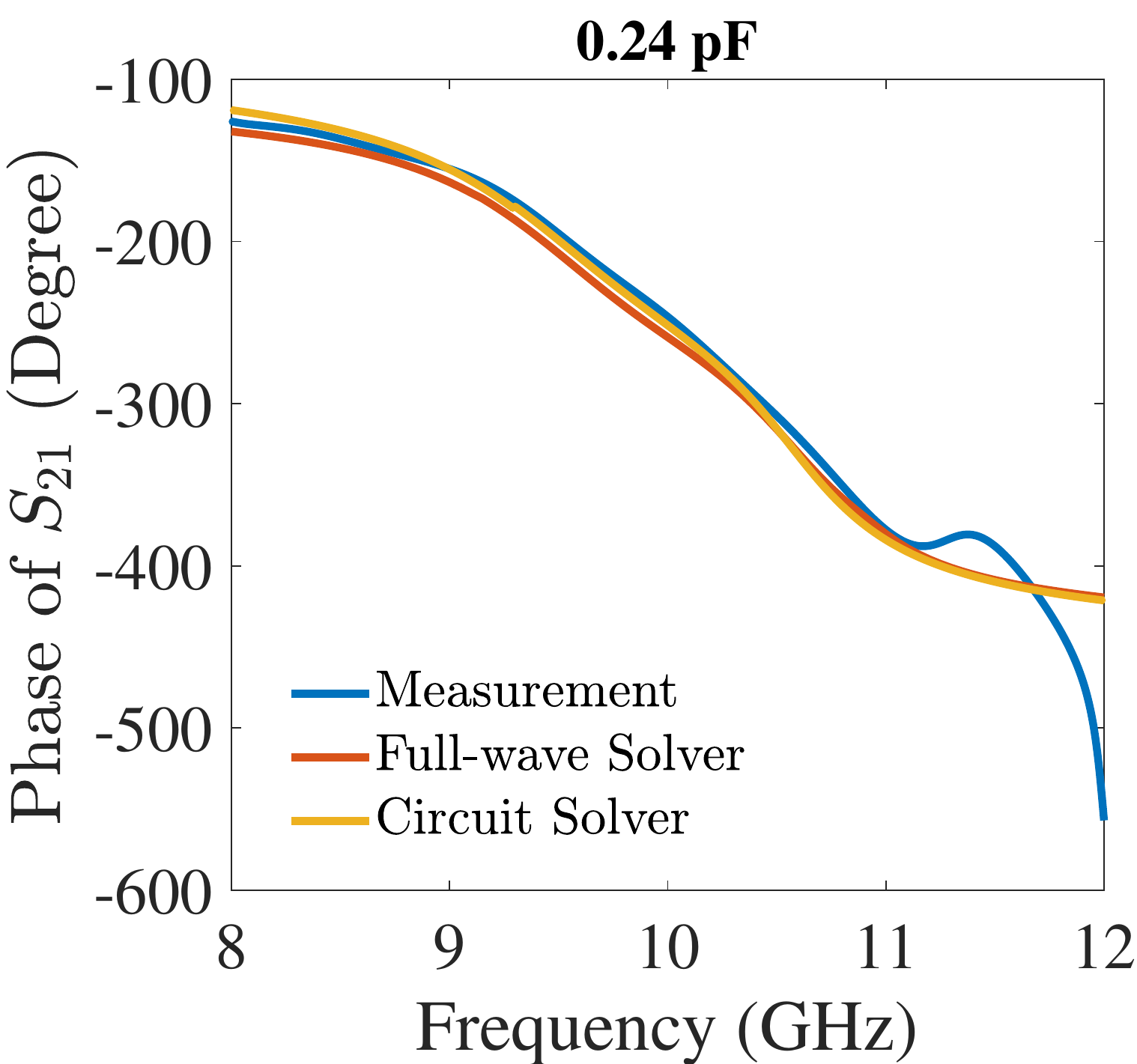}%
}\hfill
\subfloat[\label{sfig:Phase18}]{%
  \includegraphics[clip,width=0.32\columnwidth]{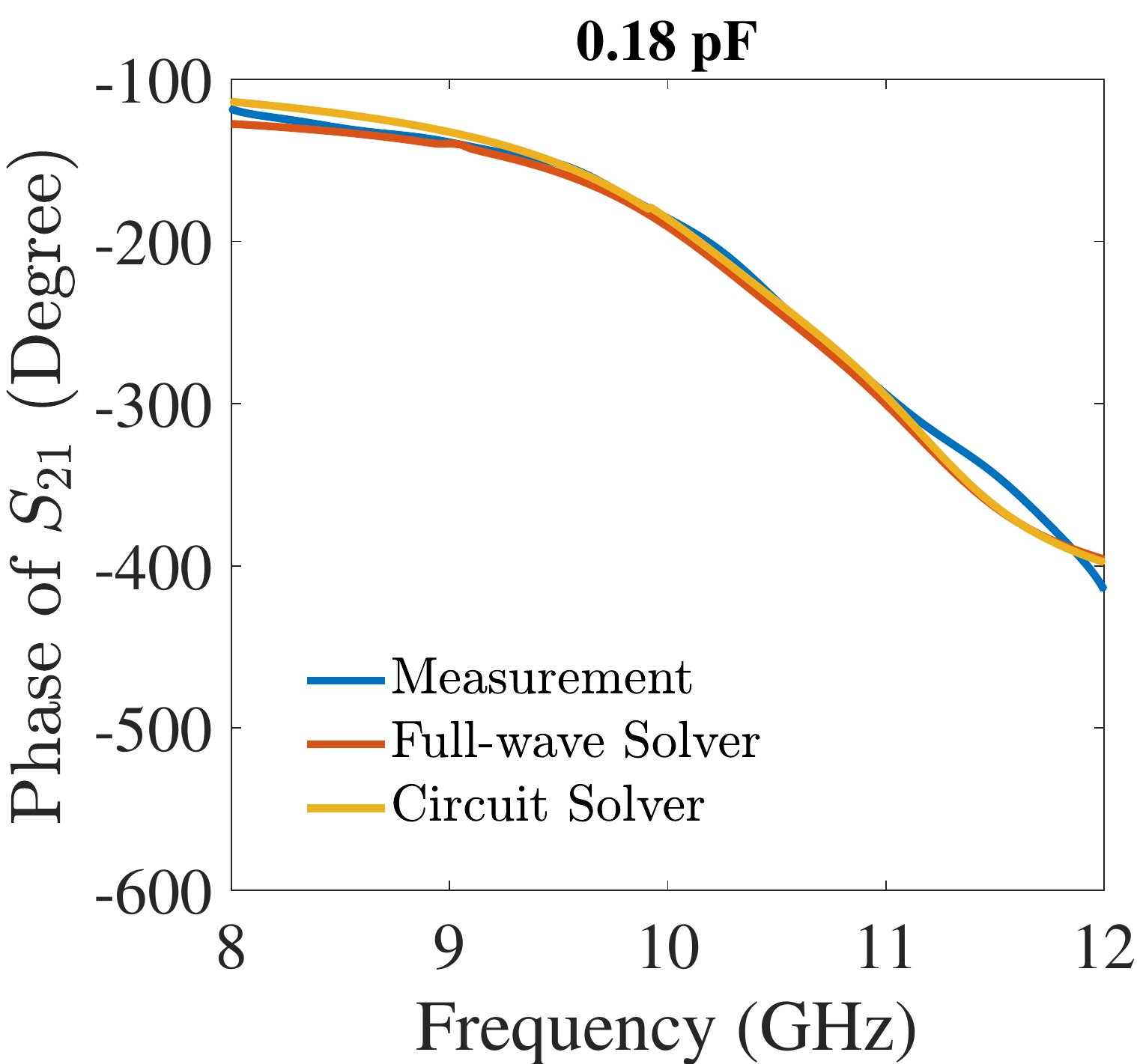}%
}\qquad
\caption{Comparison of the transmission coefficient entries of the transparent, tunable metasurface phase shifter in measurement, full-wave solver (HFSS) and circuit solver (ADS).}
\label{fig:BiasCompare}
\end{figure}

Our measurement results showed that the varactor capacitance versus reverse bias voltage characteristic quoted on the diode data sheet is inaccurate at the operating frequency of 10 GHz \cite{PhysRevX.9.011036}. An accurate capacitance vs. reverse bias voltage relationship of the varactors was extracted by comparing transmission measurements at different bias voltages to simulation results for different capacitance values, as shown in Table \ref{tab:BiasVoltage}. The simulated and measured transmission coefficient entries of the transparent, tunable metasurface phase shifter (given in Fig. 3a) are shown in Fig. \ref{fig:BiasCompare}. For three representative bias voltages 7 V, 10 V and 13 V, the measurement results are given as the blue curves in Fig. \ref{fig:BiasCompare}. The full wave simulation results under the corresponding varactor capacitances (0.345 pF, 0.24 pF and 0.18 pF respectively, as given in Table \ref{tab:BiasVoltage}) are given as the red curves. In addition, the simulation results of the circuit model of Fig. 2b under the corresponding varactor capacitances are given as the orange curves. Close agreement is observed among measurement, full-wave simulation and circuit simulation results of the phase-shifting metasurface, verifying the parameters given in Table II.

\begin{table}[t]
\caption{\label{tab:BiasVoltage}%
Measured and simulated varactor capacitance for each biasing voltage.}
\centering
\setlength{\tabcolsep}{0.5pt}
\begin{tabular}{|>{\centering\arraybackslash}p{0.3\columnwidth}| >{\centering\arraybackslash}p{0.3\columnwidth}|>{\centering\arraybackslash}p{0.33\columnwidth}|}
\hline
 Bias voltage used in ADS (V)&Varactor capacitance (pF)&Bias voltage used in measurement (V)\\
\hline
4.88&0.345&7\\
5.95&0.30&8\\
7.35&0.26&9\\
8.33 & 0.24&10\\
9.69& 0.22&11\\
11.61& 0.20&12\\
14.70& 0.18&13\\
\hline
\end{tabular}
\end{table}

\begin{figure}[t]
\centering
\includegraphics[clip,width=\columnwidth]{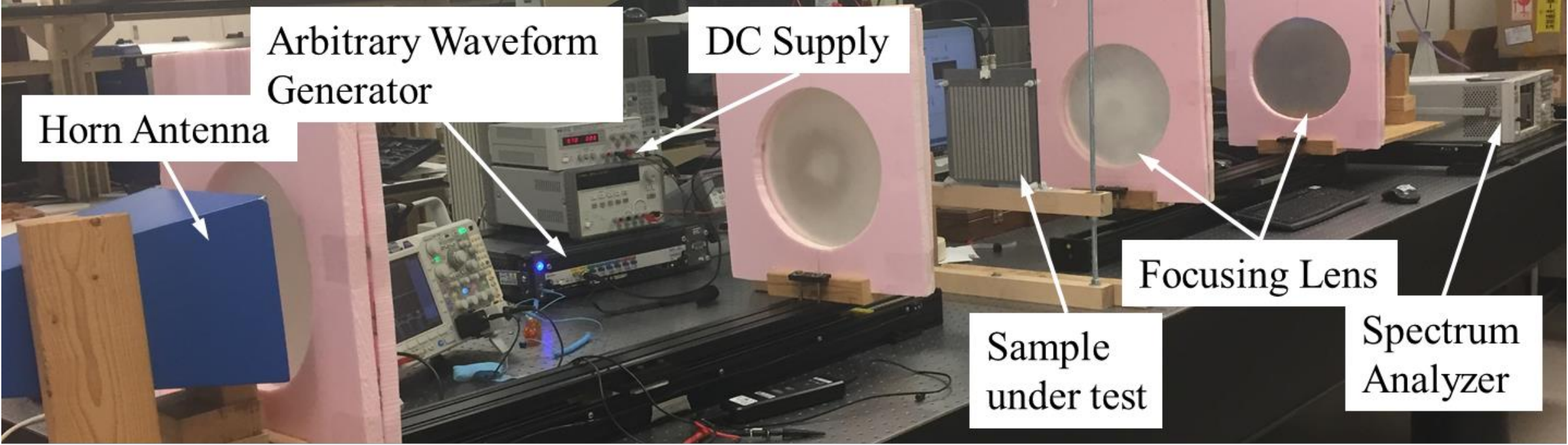}% Here is how to import EPS art
\caption{\label{fig:MeaSetup} Photograph of the quasi-optical, free-space measurement system.}
\end{figure}

\begin{figure}[b]
\centering
\subfloat[\label{sfig:MeaVol_BFS}]{%
  \includegraphics[clip,width=0.45\columnwidth]{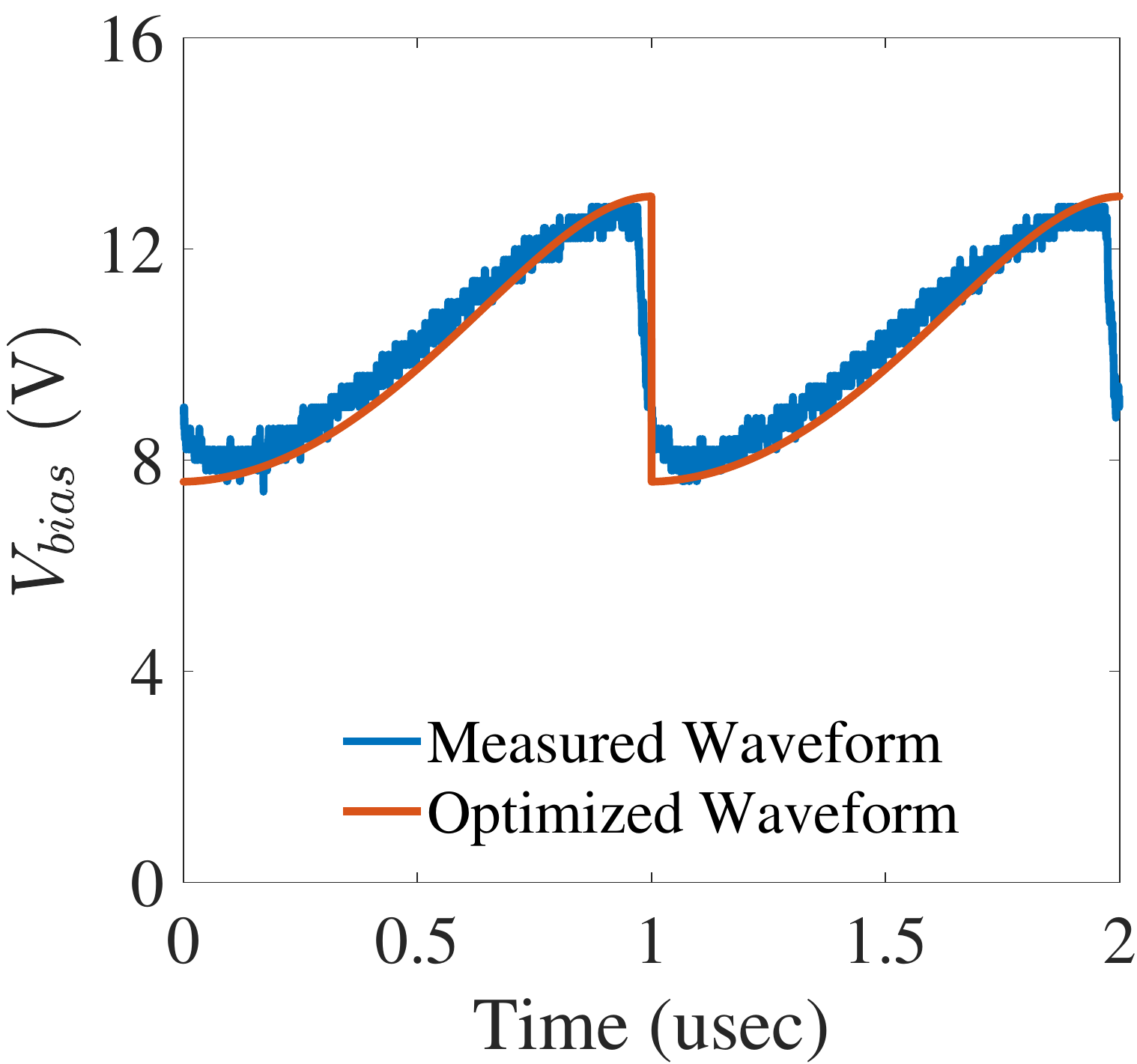}%
}\hfill
\subfloat[\label{sfig:MeaSpec_BFS}]{%
  \includegraphics[clip,width=0.45\columnwidth]{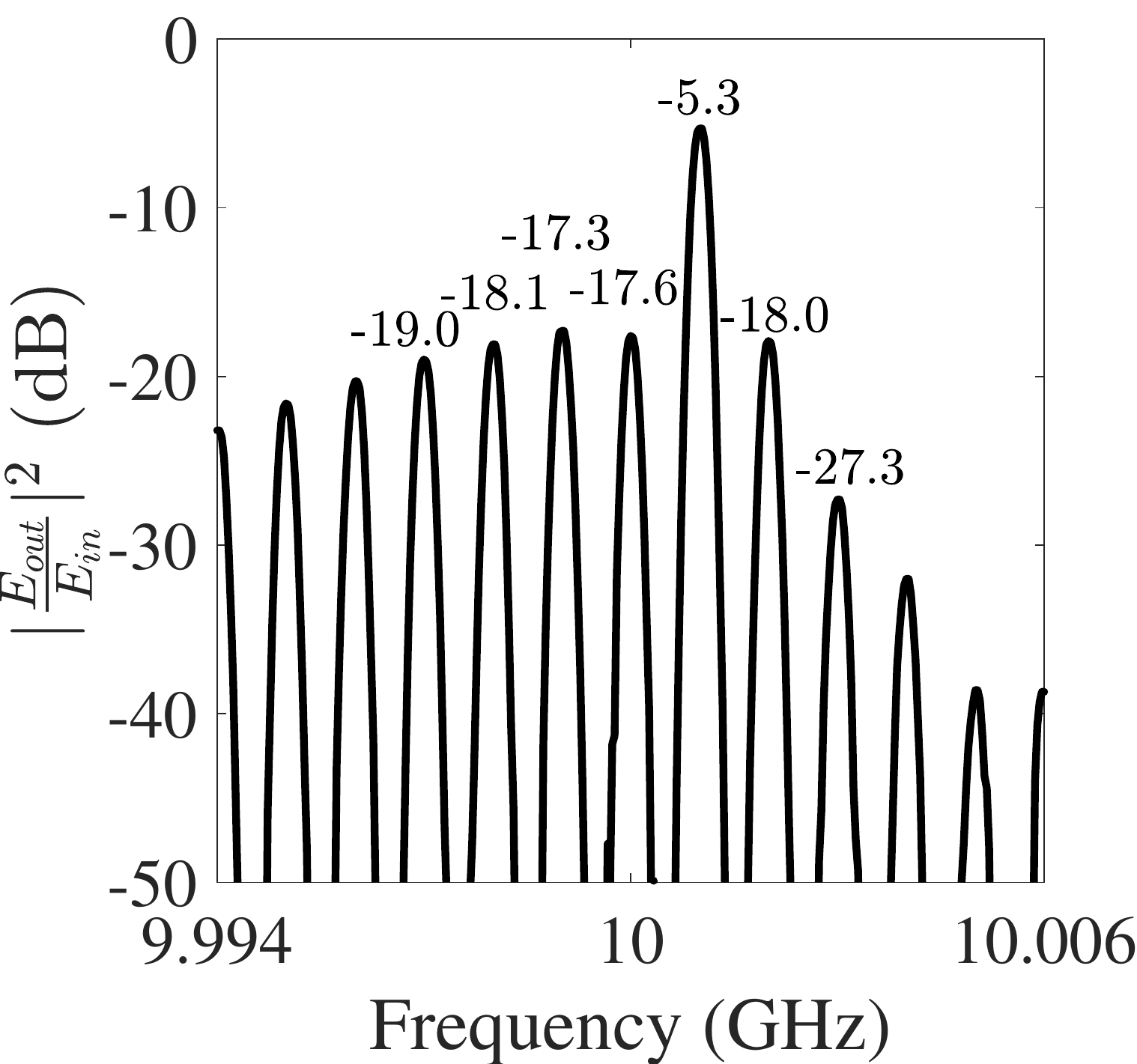}%
}
\caption{(a) Bias waveform used in measurement. (b) Measured transmission spectrum for a 10 GHz continuous wave signal incident on the transmissive frequency translator.}
\label{fig:BFSMeasurement}
\end{figure}

The fabricated frequency translator was measured under a quasi-optical Gaussian beam system, as shown in Fig. \ref{fig:MeaSetup}. A rectangular horn antenna (i.e., Dorado GH-90-25) along with a pair of lenses radiate a Gaussian beam with a beam waist of 114 mm in diameter. The focal lengths of the lenses used are 45 cm. The fabricated frequency translator had a dimension of $170\times 168$ mm$^2$ \cite{PhysRevX.9.011036}. The minimum dimension of the frequency translator is approximately $1.5$ times the beam waist, to limit diffraction. An analog signal generator (Agilent N5183) was used to feed the transmit horn antenna with a $-20$ dBm signal at 10 GHz. The amplitude of the incident wave impinging on the metasurfaces was measured to be $-26.5$ dBm. The receive antenna was connected to a spectrum analyzer (Agilent E4446A).  The path loss of the system was measured at the operating frequency and calibrated out of the measurements. The ratio of the calibrated output signal power to input signal power provided the transmission coefficient at each measured frequency harmonic. The bias waveform $V_{bias}(t)$ used in measurement was determined by applying the voltage relationship given in Table \ref{tab:BiasVoltage} to the optimized bias waveform used in simulation, as shown in  Fig. \ref{sfig:MeaVol_BFS} (red curve). The required $V_{bias}(t)$ was provided by a Keysight M8195A arbitrary waveform generator, and amplified with ZHL-3A-S(+) RF amplifiers from Mini-Circuits. The bias waveform across the diode was measured using a differential probe (Tektronix TMDP0200) and Tektronix oscilloscope MDO3024. The measured and simulated bias waveforms are compared in Fig. \ref{sfig:MeaVol_BFS}. The measured transmission spectrum of the fabricated serrodyne frequency translator is shown in Fig. \ref{sfig:MeaSpec_BFS}. A $5.3$ dB conversion loss and $12$
dB sideband suppression was achieved in experiment. The difference between the measured and simulated transmission spectrum is attributed to the performance of the biasing network. The biasing network was not optimized for a time-varying bias, but originally designed for a DC bias.

\section{Conclusion}
In conclusion, a transparent serrodyne frequency translator based on time-modulated metasurfaces is reported at X-band frequencies. The fabricated frequency translator achieves $5.3$ dB conversion loss with $12$ dB sideband suppression in experiment. The development of time-modulated, phase tunable metasurfaces opens new and exciting areas of research.  Specifically, transmissive phase-modulating metasurfaces can be cascaded together, or multiple in tandem, as in \cite{doerr2014silicon} and \cite{reiskarimian2016magnetic}, to produce non-reciprocal devices such as broadband isolators and circulators in a compact, low-profile form factor for Gaussian beam (optical and quasi-optical) systems.

\section*{Acknowledgment}
This work was supported by DSO National Laboratories under Contract No. DSOCO15027, and the Air Force Office of Scientific Research (AFOSR) under Grant No. FA9550-15-1-0101, and the AFOSR MURI program No. FA9550-18-1-0379.

\bibliographystyle{IEEEtran}
\bibliography{FrequencyConverter}

\begin{thebibliography}{10}
\providecommand{\url}[1]{#1}
\csname url@samestyle\endcsname
\providecommand{\newblock}{\relax}
\providecommand{\bibinfo}[2]{#2}
\providecommand{\BIBentrySTDinterwordspacing}{\spaceskip=0pt\relax}
\providecommand{\BIBentryALTinterwordstretchfactor}{4}
\providecommand{\BIBentryALTinterwordspacing}{\spaceskip=\fontdimen2\font plus
\BIBentryALTinterwordstretchfactor\fontdimen3\font minus
  \fontdimen4\font\relax}
\providecommand{\BIBforeignlanguage}[2]{{%
\expandafter\ifx\csname l@#1\endcsname\relax
\typeout{** WARNING: IEEEtran.bst: No hyphenation pattern has been}%
\typeout{** loaded for the language `#1'. Using the pattern for}%
\typeout{** the default language instead.}%
\else
\language=\csname l@#1\endcsname
\fi
#2}}
\providecommand{\BIBdecl}{\relax}
\BIBdecl

\bibitem{sounas2017non}
D.~L. Sounas and A.~Al{\`u}, ``Non-reciprocal photonics based on time
  modulation,'' \emph{Nature Photonics}, vol.~11, no.~12, p. 774, 2017.

\bibitem{caloz2018electromagnetic}
C.~Caloz, A.~Al{\`u}, S.~Tretyakov, D.~Sounas, K.~Achouri, and Z.-L.
  Deck-L{\'e}ger, ``Electromagnetic nonreciprocity,'' \emph{Physical Review
  Applied}, vol.~10, no.~4, p. 047001, 2018.

\bibitem{zucker1961traveling}
H.~Zucker, ``Traveling-wave parametric amplifier analysis using difference
  equations,'' \emph{Proceedings of the IRE}, vol.~49, no.~3, pp. 591--598,
  1961.

\bibitem{risk1984acousto}
W.~Risk, R.~Youngquist, G.~Kino, and H.~J. Shaw, ``Acousto-optic frequency
  shifting in birefringent fiber,'' \emph{Optics letters}, vol.~9, no.~7, pp.
  309--311, 1984.

\bibitem{cheng1992baseband}
Z.~Cheng and C.~Tsai, ``Baseband integrated acousto-optic frequency shifter,''
  \emph{Applied physics letters}, vol.~60, no.~1, pp. 12--14, 1992.

\bibitem{jaffe1965microwave}
J.~S. Jaffe and R.~Mackey, ``Microwave frequency translator,'' \emph{IEEE
  Transactions on Microwave Theory and Techniques}, vol.~13, no.~3, pp.
  371--378, 1965.

\bibitem{cumming1957serrodyne}
R.~C. Cumming, ``The serrodyne frequency translator,'' \emph{Proceedings of the
  IRE}, vol.~45, no.~2, pp. 175--186, 1957.

\bibitem{klein1967digilator}
G.~Klein and L.~Dubrowsky, ``The digilator, a new broadband microwave frequency
  translator,'' \emph{IEEE Transactions on Microwave Theory and Techniques},
  vol.~15, no.~3, pp. 172--179, 1967.

\bibitem{wong1982electro}
K.~Wong, R.~De~La~Rue, and S.~Wright, ``Electro-optic-waveguide frequency
  translator in linbo 3 fabricated by proton exchange,'' \emph{Optics letters},
  vol.~7, no.~11, pp. 546--548, 1982.

\bibitem{johnson2010broadband}
D.~Johnson, J.~Hogan, S.-W. Chiow, and M.~Kasevich, ``Broadband optical
  serrodyne frequency shifting,'' \emph{Optics letters}, vol.~35, no.~5, pp.
  745--747, 2010.

\bibitem{houtz2009wideband}
R.~Houtz, C.~Chan, and H.~M{\"u}ller, ``Wideband, efficient optical serrodyne
  frequency shifting with a phase modulator and a nonlinear transmission
  line,'' \emph{Optics Express}, vol.~17, no.~21, pp. 19\,235--19\,240, 2009.

\bibitem{lucyszyn199424}
S.~Lucyszyn, I.~D. Robertson, and H.~Aghvami, ``24 ghz serrodyne frequency
  translator using a 360/spl deg/analog cpw mmic phase shifter,'' \emph{IEEE
  Microwave and Guided Wave Letters}, vol.~4, no.~3, pp. 71--73, 1994.

\bibitem{wu2010generation}
R.~Wu, V.~Supradeepa, C.~M. Long, D.~E. Leaird, and A.~M. Weiner, ``Generation
  of very flat optical frequency combs from continuous-wave lasers using
  cascaded intensity and phase modulators driven by tailored radio frequency
  waveforms,'' \emph{Optics letters}, vol.~35, no.~19, pp. 3234--3236, 2010.

\bibitem{jessen1992generation}
P.~Jessen and M.~Kristensen, ``Generation of a frequency comb with a double
  acousto-optic modulator ring,'' \emph{Applied optics}, vol.~31, no.~24, pp.
  4911--4913, 1992.

\bibitem{cundiff2010optical}
S.~T. Cundiff and A.~M. Weiner, ``Optical arbitrary waveform generation,''
  \emph{Nature Photonics}, vol.~4, no.~11, p. 760, 2010.

\bibitem{ramaccia2017doppler}
D.~Ramaccia, D.~L. Sounas, A.~Al{\`u}, A.~Toscano, and F.~Bilotti, ``Doppler
  cloak restores invisibility to objects in relativistic motion,''
  \emph{Physical Review B}, vol.~95, no.~7, p. 075113, 2017.

\bibitem{taravati2017mixer}
S.~Taravati and C.~Caloz, ``Mixer-duplexer-antenna leaky-wave system based on
  periodic space-time modulation,'' \emph{IEEE Transactions on Antennas and
  Propagation}, vol.~65, no.~2, pp. 442--452, 2017.

\bibitem{henthorn2017bit}
S.~Henthorn, K.~L. Ford, and T.~O’Farrell, ``Bit-error-rate performance of
  quadrature modulation transmission using reconfigurable frequency selective
  surfaces,'' \emph{IEEE Antennas and Wireless Propagation Letters}, vol.~16,
  pp. 2038--2041, 2017.

\bibitem{zhang2018space}
L.~Zhang, X.~Q. Chen, S.~Liu, Q.~Zhang, J.~Zhao, J.~Y. Dai, G.~D. Bai, X.~Wan,
  Q.~Cheng, G.~Castaldi \emph{et~al.}, ``Space-time-coding digital
  metasurfaces,'' \emph{Nature communications}, vol.~9, no.~1, p. 4334, 2018.

\bibitem{zhao2018programmable}
J.~Zhao, X.~Yang, J.~Y. Dai, Q.~Cheng, X.~Li, N.~H. Qi, J.~C. Ke, G.~D. Bai,
  S.~Liu, S.~Jin \emph{et~al.}, ``Programmable time-domain digital coding
  metasurface for nonlinear harmonic manipulation and new wireless
  communication systems,'' \emph{National Science Review}, 2018.

\bibitem{bekele2013pulse}
E.~T. Bekele, L.~Poli, P.~Rocca, M.~D'Urso, and A.~Massa, ``Pulse-shaping
  strategy for time modulated arrays—analysis and design,'' \emph{IEEE
  Transactions on Antennas and Propagation}, vol.~61, no.~7, pp. 3525--3537,
  2013.

\bibitem{ghelfi2012photonic}
P.~Ghelfi, F.~Scotti, F.~Laghezza, and A.~Bogoni, ``Photonic generation of
  phase-modulated rf signals for pulse compression techniques in coherent
  radars,'' \emph{Journal of Lightwave Technology}, vol.~30, no.~11, pp.
  1638--1644, 2012.

\bibitem{shaltout2015time}
A.~Shaltout, A.~Kildishev, and V.~Shalaev, ``Time-varying metasurfaces and
  lorentz non-reciprocity,'' \emph{Optical Materials Express}, vol.~5, no.~11,
  pp. 2459--2467, 2015.

\bibitem{hadad2015space}
Y.~Hadad, D.~Sounas, and A.~Alu, ``Space-time gradient metasurfaces,''
  \emph{Physical Review B}, vol.~92, no.~10, p. 100304, 2015.

\bibitem{shi2016dynamic}
Y.~Shi and S.~Fan, ``Dynamic non-reciprocal meta-surfaces with arbitrary phase
  reconfigurability based on photonic transition in meta-atoms,'' \emph{Applied
  Physics Letters}, vol. 108, no.~2, p. 021110, 2016.

\bibitem{taravati2017nonreciprocal}
S.~Taravati, N.~Chamanara, and C.~Caloz, ``Nonreciprocal electromagnetic
  scattering from a periodically space-time modulated slab and application to a
  quasisonic isolator,'' \emph{Physical Review B}, vol.~96, no.~16, p. 165144,
  2017.

\bibitem{estep2014magnetic}
N.~A. Estep, D.~L. Sounas, J.~Soric, and A.~Al{\`u}, ``Magnetic-free
  non-reciprocity and isolation based on parametrically modulated
  coupled-resonator loops,'' \emph{Nature Physics}, vol.~10, no.~12, p. 923,
  2014.

\bibitem{sounas2013giant}
D.~L. Sounas, C.~Caloz, and A.~Alu, ``Giant non-reciprocity at the
  subwavelength scale using angular momentum-biased metamaterials,''
  \emph{Nature communications}, vol.~4, p. 2407, 2013.

\bibitem{qin2014nonreciprocal}
S.~Qin, Q.~Xu, and Y.~E. Wang, ``Nonreciprocal components with distributedly
  modulated capacitors,'' \emph{IEEE Transactions on Microwave Theory and
  Techniques}, vol.~62, no.~10, pp. 2260--2272, 2014.

\bibitem{ramaccia2015nonreciprocal}
D.~Ramaccia, D.~L. Sounas, A.~Al{\`u}, F.~Bilotti, and A.~Toscano,
  ``Nonreciprocal horn antennas using angular momentum-biased metamaterial
  inclusions,'' \emph{IEEE Transactions on Antennas and Propagation}, vol.~63,
  no.~12, pp. 5593--5600, 2015.

\bibitem{PhysRevX.9.011036}
\BIBentryALTinterwordspacing
Z.~Wu, Y.~Ra'di, and A.~Grbic, ``Tunable metasurfaces: A polarization rotator
  design,'' \emph{Phys. Rev. X}, vol.~9, p. 011036, Feb 2019. [Online].
  Available: \url{https://link.aps.org/doi/10.1103/PhysRevX.9.011036}
\BIBentrySTDinterwordspacing

\bibitem{pozar2009microwave}
D.~M. Pozar, \emph{Microwave engineering}.\hskip 1em plus 0.5em minus
  0.4em\relax John Wiley \& Sons, 2009.

\bibitem{macom}
M.~T. Solutions, \emph{Solderable GaAs Constant Gamma Flip-Chip Varactor Diode,
  MAVR-000120-1411}, available at
  \url{https://www.macom.com/products/product-detail/MAVR-000120-14110P}.

\bibitem{doerr2014silicon}
C.~Doerr, L.~Chen, and D.~Vermeulen, ``Silicon photonics broadband
  modulation-based isolator,'' \emph{Optics express}, vol.~22, no.~4, pp.
  4493--4498, 2014.

\bibitem{reiskarimian2016magnetic}
N.~Reiskarimian and H.~Krishnaswamy, ``Magnetic-free non-reciprocity based on
  staggered commutation,'' \emph{Nature communications}, vol.~7, p. 11217,
  2016.

\end{thebibliography}

\end{document}